\documentstyle[12pt,jeep]{article}
\numberbysection
\date{\today}
\begin{document}
\title{\bf Parametrically Excited Hamiltonian Partial Differential Equations}
\author{
  E. Kirr \thanks{
 Department of Mathematics, University of Michigan, Ann Arbor, 
  MI and Department of Applied Mathematics, University "Babe\c s-Bolyai", Cluj, Romania} \hspace{.05 in}
 and
  M.I. Weinstein \thanks{
 Mathematical Sciences Research, Bell Laboratories - Lucent Technologies,
  Murray Hill, NJ 
 }
}
   \baselineskip=18pt
   \maketitle
\def\M{(\varepsilon^2\Gamma)^{-\alpha}}
\def\MM{(\varepsilon^2\Gamma/2)^{-\alpha}}
\def\un{\underline}
\def\nn{\nonumber}
\newcommand{\no}{\nonumber}
\newcommand{\be}{\begin{equation}}
\newcommand{\ee}{\end{equation}}
\newcommand{\ba}{\begin{eqnarray}}
\newcommand{\ea}{\end{eqnarray}}
\newcommand{\ve}{\varepsilon}
\newcommand{\nit}\noindent
\newcommand{\D}\partial
\def\Pc{{\bf P_c}}
\def\tA{{\tilde A}}
\newcommand{\lan}\langle
\newcommand{\ran}\rangle
\newcommand{\wpl}{w_+}
\newcommand{\wmi}{w_-}
\newcommand{\spectrum}{\sigma_{\rm cont}(H_0)}
\newcommand{\la}{\lambda}
\newcommand{\ra}{\rightarrow}
\newcommand{\Z}{{\rm Z\kern-.35em Z}}
\newcommand{\N}{{\rm I\kern-.15em N}}
\newcommand{\res}{I_{res}}
\newcommand{\bP}{{\rm I\kern-.15em P}}
\newcommand{\Q}{\kern.3em\rule{.07em}{.65em}\kern-.3em{\rm Q}}
\newcommand{\R}{{\rm I\kern-.15em R}}
\newcommand{\h}{{\rm I\kern-.15em H}}
\newcommand{\C}{\kern.3em\rule{.07em}{.55em}\kern-.3em{\rm C}}
\newcommand{\T}{{\rm T\kern-.35em T}}
\newtheorem{theo}{Theorem}[section]
\newtheorem{defin}{Definition}[section]
\newtheorem{prop}{Proposition}[section]
\newtheorem{lem}{Lemma}[section]
\newtheorem{cor}{Corollary}[section]
\newtheorem{rmk}{Remark}[section]
\begin{abstract}
\nit
Consider a linear autonomous Hamiltonian system with a time periodic bound state solution.
In this paper we study the structural instability of this bound state 
relative to time almost periodic perturbations which are 
 small, localized and  Hamiltonian.
  This class of 
perturbations includes those whose time dependence is 
 periodic, but encompasses a large class of those with finite (quasiperiodic)
 or infinitely many  non-commensurate frequencies. Problems of the type 
considered arise in many areas of application  including ionization physics
and the propagation of light in optical fibers in the presence of defects.
 The mechanism of instability is radiation  damping due to 
  resonant coupling of the bound state to the continuum modes by the
 time-dependent
  perturbation. This results in a transfer of energy from the discrete
 modes to the continuum.
  The rate of decay of solutions is
 slow and hence the decaying bound states can be viewed as metastable. 
 These 
 results generalize those of 
 A. Soffer and M.I. Weinstein, who treated localized time-periodic 
 perturbations of a particular form. 
 In the present work, 
 new analytical issues need to be addressed in view of (i) the presence of 
 infinitely many frequencies which may resonate with the continuum as well as 
 (ii) the possible accumulation of such resonances in the continuous spectrum.
 The theory is applied to a general class of Schr\"odinger operators.
\end{abstract}
\thispagestyle{empty}
\vfil\eject
 \section{Introduction}
\medskip

\subsection{Overview}
\medskip

Consider a dynamical system of the form:
\be
i\D_t\phi\ =\ H_0\ \phi,
\label{eq:unperturbed}\ee
where $H_0$ denotes a self-adjoint operator on a Hilbert space
 ${\cal H}$. We further assume
that $H_0$ has only one eigenstate $\psi_0\in{\cal H}$ with corresponding
simple eigenvalue $\lambda_0$. Thus, 
\be b_*(t)\ =\ e^{-i\lambda_0 t}\psi_0\label{eq:bb}\ee
 is a time-periodic
 {\it bound state} solution of the dynamical system (\ref{eq:unperturbed}). 
We next introduce the perturbed dynamical system:
\be
i\D_t\phi\ =\ \left(\ H_0\ +\ \ve W(t)\ \right)\phi.
\label{eq:perturbed}\ee
  In this paper we prove that if the perturbation  
, $\ve W(t)$,  is a small, "generic''  and  
 almost periodic in time 
\footnote{See the appendix in section \ref{se:app} as well as  \cite{kn:Bohr},
 \cite{kn:Levitan} for definitions and results on
almost periodic functions.}, 
 then solutions
 of the perturbed dynamical system (\ref{eq:perturbed}) tend to zero
 as $t\to\pm\infty$. It follows that the state, $b_*(t)$,  does not 
continue or deform to a time periodic or even time almost periodic state. 
Thus, $b_*(t)$ is structurally unstable with respect to this 
 class of perturbations.
  Our methods yield a detailed description of the 
  transient ($t$ large but finite)  and long time ($t\to\pm\infty$)
  behavior solutions to the initial value problem. 
 Theorems \ref{th:twofive}-\ref{th:twotwo} contain precise statements of our main results.
  The following picture emerges concerning time evolution  
 (\ref{eq:perturbed}) for initial data given by the bound state, $\psi_0$,
  of the unperturbed problem. Let 
\be P(t)= |(\ \psi_0,\phi(t)\ )|^2,\label{eq:Pdef}\ee
the modulus square
of the projection of the solution at time $t$ onto the state $\psi_0$ 
\footnote{ $(f,g)$ denotes the inner product of $f$ and $g$. If $\psi_0$ is
normalized then $P(t)$ has the 
 quantum mechanical interpretation of the probability 
 that the system at time $t$ is in the state $\psi_0$.}. Then,

\noindent (i)  $P(t)\ \sim\ 1-C_W\ |t|^2$, for $|t|$ small,\footnote{ We
do not discuss the short time behavior in this article; see \cite{kn:MSW}.
This small time behavior is related to the "watched pot'' effect in 
 quantum measurement theory \cite{kn:Newton}.}

\noindent (ii)  $P(t)\ \sim\ \exp(-2\ve^2\Gamma t)$ for 
 $t \le  {\cal O}((\ve^2\Gamma)^{-1}), \ 
 \Gamma\ = {\cal O}( W^2 )$, and 

\noindent (iii)  $P(t)\ \sim \ \langle t\rangle^{-\alpha}$ for 
$|t|>>(\ve^2\Gamma)^{-1}$,
 for some $\alpha>0$.
 
The time $\tau =(\ve^2\Gamma)^{-1}$ is called the {\it lifetime} of the state $b_*(t)$,
which can be thought of as being {\it metastable} due to its slow decay.
 The mechanism for large time
  decay is resonant coupling of the bound state
with continuous spectrum due to the time-dependent perturbation.
 Our analysis makes explicit the slow transfer of energy from the 
 discrete to 
 continuum modes, and the accompanying radiation of energy out of any 
 compact set.

Phenomena of the type considered here are of importance in many areas of 
theoretical physics and applications.  Examples include:
 ionization physics \cite{kn:CDG, kn:G-P, kn:L-L} and 
  in the propagation of light in optical fibers in the
  presence of defects \cite{kn:Marcuse}; see the discussion below. 

The results of this article
  generalize those of Soffer and Weinstein \cite{kn:nh}, where
the case:
\be W(t)\ =\ \cos(\mu t)\ \beta,\ \ \beta=\beta^*\no
\ee
was considered. The method used is a time dependent / dynamical systems 
 approach 
  introduced in \cite{kn:TDRT0},
\cite{kn:TDRT1} for the problem of perturbations of
operators with embedded eigenvalues in their continuous spectra,
 \cite{kn:rdamping}, in the context of resonant radiation damping of
 nonlinear systems,
as well as in \cite{kn:nh}; 
 see also \cite{kn:MSW}.
 New analytical questions must be addressed in view of (i) the presence
of infinitely many frequencies which may resonate with the continuum as well
as (ii) the possible accumulation of such resonances in the continuous 
 spectrum.  This leads to a careful use of almost periodic properties of the 
perturbation (Theorems \ref{th:twofive} and \ref{th:twosix})
 and hypothesis {\bf (H6)} (Theorem \ref{th:twotwo}), which is easily seen
to hold when the perturbation, $W(t)$, consists of a sum over finite number of 
frequencies, $\mu_j$.

A special case for which the hypotheses of our theorems are verified is the case of the  
 Schr\"odinger
operator $H_0\ =\ -\Delta\ +\ V(x)$. Here, $V(x)$ is a real-valued function of 
 $x\in\R^3$ which decays sufficiently
rapidly as $|x|\to\infty$.
In this setting Soffer and Weinstein \cite{kn:nh}
  studied in detail the structural
instability of $b_*(t)$ by considering the 
 perturbed dynamical system (\ref{eq:perturbed}), with
  $W(x,t)\ =\ \beta(x)\ \cos(\mu t)$. Here, we consider a class
 of  perturbations
 of the form $W(x,t)\ =\ \sum_j\ \beta_j(x)\ \cos{\mu_j t}$, where the
sum may be finite or infinite and where the frequencies $\mu_j$ need not be
commensurate, {\it e.g.} $W(x,t)\ =\ \beta_1(x)\ \cos{t}
\ +\ \beta_2(x)\ \cos{\sqrt{2} t}$, where $\beta_i(x),\ i=1,2$ is rapidly 
decaying as $x\to\infty$.

In addition to the problem of ionization by general time varying fields, we mention other motivations for considering the  class of time 
dependent perturbations sketched above and defined in detail in section \ref{se:mainresults}. 

\noindent (a) An area of application to which our analysis applies is 
the propagation of light through an optical fiber \cite{kn:Marcuse}. 
 In the regime where 
 backscattering can be neglected, the propagation of waves 
 down the length of the 
fiber is governed by a Schr\"odinger equation:
\be
i\D_z\phi\ =\ \left(\ -\Delta_\perp\ +\ V(x_\perp)\ \right)\phi\ +\ 
 W(x_\perp,z)\phi.
\label{eq:fibereqn}\ee
Here, $\phi$ denotes the slowly varying envelope of the highly oscillatory 
electric field, a function of $z$, the direction of propagation 
 along the fiber,  and $x_\perp\in\R^2$,
 the transverse variables. $V(x_\perp)$ denotes an unperturbed index of refraction profile
 and $W(x_\perp,z)$, the small 
  fluctuations in refractive index along the fiber.  
These can 
  arise due to defects introduced either accidentally or by 
 design.  The models considered allow for distributions of defects which are far more general than periodic. Our analysis addresses the simple situation of energy in a single
transverse mode propagating and being radiated away due to coupling by defects to continuum  
modes. The bound state channel sees an effective damping. In particular the results of this paper have been applied to a study of structural instability of so-called breather modes of planar "soliton wave guides" \cite{kn:MSW}.
The case of multiple transverse modes is of great interest \cite{kn:Marcuse}.
  Here one has
 the  phenomena of coupling among discrete modes as well as 
the coupling of discrete to continuum / radiation modes \cite{kn:KW}. There is 
 extensive interesting work on 
this problem in the case where $W(x_\perp,z)$ is a stochastic process in $z$
 and radiation is neglected \cite{kn:kp}.

\noindent (b) Nonlinear problems can be viewed as linear time-dependent
  potential problems 
where the time-dependent potential is given by the solution. 
 {\it A priori} one knows little  
about the time dependence of the solution of a nonlinear problem. Nonlinearity 
is expected, in general, to excite infinitely many frequencies. Therefore 
  results of a general 
nature for potentials with very general time dependence are of interest.
This point of view is adopted by I.M. Sigal \cite{kn:sigalcmp}, 
 \cite{kn:sigaljapan}, who considers the case
where the nonlinear term defines a time-periodic perturbation, and then proceeds to study the resonance problem via time-independent Floquet analysis applied to the so-called Floquet Hamiltonian. The dilation analytic techniques used were first applied in the context of time-periodic Hamiltonians by  Yajima \cite{kn:Ya, kn:Yaj-stark, kn:Yaj-chargetransfer}. Floquet type methods were also used in the time-periodic context by Vainberg \cite{kn:Vain}.
The general class of perturbations we consider are not treatable by Floquet analysis and time-dependent analysis appears necessary. 
\medskip

\subsection{Outline of the method}
\medskip

We now give a brief outline of our approach. 
  For simplicity consider the 
initial value problem:
\ba
i\D_t\phi(t,x)\ &=&\ H_0\ \phi(t,x)\ +\ 
  \ve W(t,x)
\ \phi(t,x), \label{eq:modelprob}\\
\phi|_{t=0}\ &=&\ \phi(0)\label{eq:phidata}
\ea
where  
\be H_0=-\Delta\ +V(x),\ 
    W(t,x)=g(t)\ \beta(x),\ 
    g(t)=\sum_j\ g_j\ e^{-i\mu_j t}\label{eq:gdef}\ee
is a real-valued  almost periodic function of $t$, and 
 $\beta(x)$ is a real-valued and rapidly decaying function of $x$
 as $|x|\to\infty$.
The unperturbed problem ($\ve=0$) 
 can be trivally written as two 
decoupled equations governing the bound state amplitude, $a(t)$, and
  dispersive components, $\phi_d(t)$,  of the solution. Specifically, let 
\be \phi(t)\ =\ a(t)\ \psi_0(x)\ +\ \phi_d(t,x),\ 
 \left(\ \psi_0,\phi_d(t)\ \right)\ =\ 0.\label{eq:decomposition}\ee
Then,
\ba
i\D_t a(t)\ =\ \lambda_0 a(t),\nonumber\\
i\D_t\ \phi_d(t,x)\ =\ H_0\ \phi_d(t,x),\label{eq:aphideps0}
\ea
with initial conditions:
\ba
a(0)\ =\ \left( \psi_0, \phi(0) \right),\nonumber\\
\phi_d(0)\ =\ \Pc\phi(0),\label{eq:aphiddata} 
\ea
where
$$\Pc f\equiv\ f-\left(\psi_0,f\right)\psi_0$$
defines the projection onto the continuous spectral part of $H_0$.

For initial data $a(0)=1$, $\phi_d(0)=0$, we have 
 $a(t)=e^{-i\lambda_0t}$,
 $\phi_d(t)\equiv 0$, corresponding to the bound state, $b_*(t)$.

We now ask:

\noindent {\it (a) Under the small perturbation $\ve W(t,x)$ 
 does the bound state deform or continue to a nearby periodic or even 
 almost periodic solution?},

\noindent {\it (b) How do solutions to the perturbed 
 initial value problem behave as $|t|\to\infty$}?

For small perturbations $\ve W(t,x)$ it is natural to use the decomposition
 (\ref{eq:decomposition}). Substitution of (\ref{eq:decomposition}) into 
 (\ref{eq:perturbed}) yields a weakly coupled system for 
$a(t)$ and $\phi_d(t)$.  This system is derived and analyzed in detail in 
sections 4--6. 

In order to illustrate the main idea, we introduce a simplified 
system having the same general character:
\ba
i\D_t a(t) \ &=&\ 
 \lambda_0\ a(t)\ +\  \ve g(t)\ \left( \beta\psi_0,\phi_d(t) \right)  
\nonumber\\
i\D_t\phi_d(t,x)\ &=&\ -\Delta\phi_d(t,x)\ +\ 
 \ve a(t)g(t)\beta(x)\psi_0(x).
\label{eq:modelsystem}\ea
Here, we have replaced $H_0$ on its continuous spectral part
 by $-\Delta$.

If $\ve \beta$ is small then $A(t)\equiv e^{i\lambda_0 t}a(t)$ is slowly varying
 ($\D_tA(t)={\cal O}(\ve \beta)$). In particular, we have
\ba
i\D_t A(t) \ &=&\ 
 \ve e^{i\lambda_0 t}\ g(t)\ \left(\ \beta\psi_0,\phi_d(t)\ \right) 
\nonumber\\
i\D_t\phi_d(t,x)\ &=&\ -\Delta\phi_d(t,x)\ + 
                       A(t) e^{-i\lambda_0 t}\ \ve g(t)\beta(x)\psi_0(x).
\label{eq:modelsystem1}\ea
Viewing $A(t)$ as nearly constant, we see that the inhomogeneous source term in 
 (\ref{eq:modelsystem1}) has frequencies $\lambda_0+\mu_j$;
  see (\ref{eq:gdef}). Therefore, if $\lambda_0+\mu_j>0$, for some $j$
 then $\lambda_0+\mu_j$ lies in the continuous spectrum of $-\Delta$ ($H_0$)
  and therefore
 $\phi_d$ satisfies a resonantly forced wave equation. A careful expansion and 
analysis to 
 second order in the perturbation $\ve W(t)$ 
 (see the proof of Proposition 4.1)
 reveals the system for  $A(t)$ and $\phi_d(t)$ can be rewritten in the following form, in which the effect of this resonance is made explicit:
\ba
\D_tA(t)\ &=&\ ( -\ve^2\Gamma\ +\ \rho(t)\ )\ A(t)\ +\  E(t;A(t),\phi_d(t)).
 \label{eq:nfA}\\
i\D_t\phi_d(t,x)\ &=&\  H_0\ \phi_d(t,x)\ +\ \Pc\ F(t,x;A(t),\phi_d(t)).
\label{eq:nfphid}\ea
 The terms $E(t)$ and $F(t,x)$ formally tend to zero if  $A(t)$ tends to zero
and if  the "local energy'' of  $\phi_d(t)$  tends to zero as $t\to\infty$.
The strategy of sections 5 and 6 is to derive coupled estimates for $A(t)$ and
a measure of the local energy of $\phi_d$ from which one can conclude, for 
$\ve W(t)$ small,  that  solutions to (\ref{eq:nfA}-\ref{eq:nfphid}) decay in an 
appropriate sense.
The key to the decay of solutions is the constant  $\Gamma$, given by 
\be
\Gamma\ \equiv\ {\pi\over4}\ \sum_{\{j\ :\ \la_0+\mu_j>0\}}\ |g_j|^2\ 
  \left(\Pc \beta \psi_0, \; \delta(H_0 - \la_0 - \mu_j)\Pc \beta \psi_0
\right); 
 \label{eq:tdfgr1}
 \ee
see also hypothesis {\bf (H5)} of section \ref{se:mainresults}.
The quantity $\Gamma$ is a generalization of the well known Fermi golden rule 
 arising in the theory of radiative transitions in  quantum mechanics 
\cite{kn:CDG, kn:G-P, kn:L-L}. For the example at hand, (\ref{eq:gdef}),
  the sum in (\ref{eq:tdfgr1}) is 
 over all $j$ for which $\mu_j + \lambda_0$ is strictly positive, {\it i.e.}
  lies in the continuous spectrum of $H_0$.
 Thinking of $H_0$ as having a spectral decomposition in terms of 
eigenfunctions and generalized eigenfunctions,  let  $e(\lambda)$ denote 
 a generalized eigenfunction associated with the energy
$\lambda$. Then each term in the sum
 (\ref{eq:tdfgr1}) is of the form:
\be \left| \left(\ e(\lambda_0+\mu_j), \beta\psi_0\ \right)\right|^2,
\label{eq:obstructions}\ee
Thus clearly $\Gamma>0$,  generically. 

Neglecting for the moment the 
oscillatory function $\rho(t)$ in (\ref{eq:nfA}), we see that 
coupling of the bound state by the time dependent perturbation to the 
 continuum- - radiation modes, at the frequencies $\mu_j +\lambda_0>0$, leads
  to decay of the bound state.  
 The leading order of equation (\ref{eq:nfA}-\ref{eq:nfphid})
  is normal form
in which this internal damping effect is made explicit; energy is transferred
from the discrete to the continuous spectral components of the solution while 
the total energy remains independent of time:
\ba
\|\ \phi(t)\ \|^2_2\ &=&\ |a(t)|^2\ +\ \|\ \phi_d(t)\ \|_2^2\no\\
              &=&\ |a(0)|^2\ +\ \|\ \phi_d(0)\ \|_2^2.\label{eq:L2norm}
\ea

%
%
\medskip

\subsection{Energy flow; contrast with the analysis of \cite{kn:nh}}
\medskip

The goal is to show that energy flows out of the bound state channel into  
dispersive spectral components. The normal form above is the system in which this energy flow is made explicit.  
Once the normal form  (\ref{eq:nfA}-\ref{eq:nfphid})
  has been derived, it is natural to  
seek coupled estimates for $A(t)$ and $\phi_d(t)$ from which their decay can be
 deduced. This is implemented in section \ref{se:localdecay}. A natural first step is to introduce the auxiliary function:
\be
\tA(t)\ \equiv\ e^{\int_0^t\ \rho(s)\ ds}\ A(t),\label{eq:tA}
\ee 
for then $\tA(t)$ satisfies simplified equation of the form:
\be
\D_t\tA(t)\ =\ -\ve^2\Gamma\ \tA(t)\ +\  {\tilde E}(t;\tA(t),\phi_d(t))
\ee
If $\Re \int_0^t\ \rho(s)\ ds$ is uniformly bounded then, modulo time-decay 
 estimates on
$\tilde E(t;\tA,\phi_d)$ and $F(t;\tA,\phi_d)$, the decay of $\tA(t)$ 
 and therefore
of $A(t)$ follows. For the class of perturbations considered in  \cite{kn:nh}
 $\rho(t)$ is a periodic function, having only a finite number of 
 commensurate 
 frequencies, none of them zero. Therefore, in this case $\Re \int_0^t\ \rho(s)\ ds$
is uniformly bounded. However, in the present case $\rho(t)$ is almost periodic 
 with mean $M(\Re\rho)=0$ (see section \ref{se:app}); $\rho(t)$ is displayed in 
(\ref{eq:rhoexpansion}).  
 $\Re \rho(t)$ has, in general, 
  infinitely many frequencies, $\mu_k-\mu_j,\ k\ne j$
  which may accumulate at zero. 
 Most delicate is the 
case where, along some subsequence, $\mu_k - \mu_j\to0$. 
 It is well known that 
the integral of an almost periodic function of mean zero is not 
 necessarily bounded \cite{kn:Bohr}, so we are in need of a strategy for 
estimating the effects of $\Re \int_0^t\ \rho(s)\ ds$.  
 We address the estimation of $\Re \int_0^t\ \rho(s)\ ds$ in two
different ways corresponding to Theorem \ref{th:twotwo}
(section \ref{sse:A}) and Theorems \ref{th:twofive}-\ref{th:twosix} (section \ref{sse:B}). In section \ref{sse:A} 
 $\Re \int_0^t\ \rho(s)\ ds$ is estimated under the hypothesis {\bf (H6)} which 
 requires that the rate of accumulation of a subset of frequencies $\{\mu_j\}_{j\in I}$ is balanced by the decay of the Fourier coefficients $g_j$ as
 $j\to\infty,\ j\in I$. This leads to a bound on $\Re \int_0^t\ \rho(s)\ ds$
 (Proposition \ref{le:rhobounds}).
 In section \ref{sse:B} the estimates are based on a more refined analysis;   
 the almost periodic function $\rho(t)$, is decomposed into a part with bounded
 integral and a part which has mean zero. The latter is controlled using   
 results 
on the rate at which an almost periodic function approaches its mean.
\medskip

\subsection{Fermi golden rule and obstructions to  Poincar\'e continuation}
\medskip

In the theory or ordinary differential equations it is a standard procedure,
given a periodic solution of an unperturbed problem, to seek a periodic or 
almost periodic solution of a slightly perturbed dynamical system. We now 
investigate this procedure in the context of (\ref{eq:modelprob}) and its 
solution $b_*(t)$ for $\ve=0$.  Seek a  solution of the form:
\be
\phi(t)\ =\ b_*(t)\ +\ \phi_1(t)\ +\ {\cal O}(\ve^2\beta^2).\label{eq:pansatz}
\ee
Here, $\phi_1\ =\ {\cal O}(\ve\beta)$.\footnote{This argument is heuristic so we 
do not specify the norm with which the size of $\beta$ is measured.}
Substitution of (\ref{eq:pansatz}) into (\ref{eq:modelprob}) yields the 
equation:
\be i\D_t\phi_1\ =\ H_0\phi_1\ +\ \ve\beta\ g(t)\ b_*(t)
 .\label{eq:phi1eqn}\ee
 This equation has a solution in the class of almost periodic solutions of $t$ 
with values in the Hilbert space ${\cal H}$ only if $\beta\  g(t)\ b_*(t)$ is
"orthogonal" to the null space of $i\D_t-H_0$. 

We now derive this condition. Let $e(\zeta)$  be a solution of 
 $H_0 e(\zeta)=\zeta e(\zeta)$. Then, taking the scalar product of 
 (\ref{eq:phi1eqn}) with 
$e^{-i\zeta t}\ e(\zeta)$  and applying the operator 
$\lim_{T\uparrow\infty}\ T^{-1}\int_0^T\ \cdot\ dt$ to the 
 resulting equation gives:
\be
0\ =\ \lim_{T\uparrow\infty}\ T^{-1}\int_0^T\ e^{i\zeta t}\ e^{-i\lambda_0t}\ 
 g(t)\ dt\ \left(\ e(\zeta),\beta\psi_0\ \right).\no\ee
Substitution of the expansion for $g(t)$ yields:
\be
\sum_{j\in\Z}\ g_j\ \delta(\zeta,\lambda_0+\mu_j)\ 
\left(\ e(\zeta),\beta\psi_0\ \right)\ =\ 0,\no\ee 
where $\delta(a,b)=0$ if $a\ne b$ and $\delta(a,a)=1$. 
 If $\zeta$ , which lies in the spectrum of $H_0$, 
 satisfies $\zeta=\lambda_0+\mu_k$ for some $k\in\Z$
(which will be the case in our example if $\lambda_0+\mu_k>0$), then we  
have that: 
\be
\left(\ e(\lambda_0+\mu_k), \beta\psi_0\ \right)\ =\ 0
 \label{eq:orthogonal}
\ee
is a necessary condition for the existence of a family of solutions of 
(\ref{eq:modelprob}) which converges to $b_*(t)$ as the perturbation $W(t)$ 
tends to zero.  We immediately recognize the inner product in 
 (\ref{eq:orthogonal}) as the projection of  
 $\beta\psi_0$ onto the generalized eigenmode at the resonant frequency
 $\lambda_0+\mu_k$, which arises in (\ref{eq:tdfgr1});  see 
 also (\ref{eq:obstructions}).
 Therefore the obstruction to continuation of $b_*(t)$ to 
a nearby almost periodic state of the system can be identified with the
  damping mechanism. 

\medskip
\subsection{Outline}
\medskip

The paper is structured as follows. In section \ref{se:mainresults} we give a general 
formulation of the problem.  The hypotheses on $H_0$, the unperturbed 
Hamiltonian and $W(t)$, the perturbation are introduced and discussed. There
are two types of theorems: Theorems \ref{th:twofive} \& \ref{th:twosix} and Theorem \ref{th:twotwo}. 
 Although
the conclusions of these are quite similar, as discussed above, 
they differ in a key hypothesis 
on the perturbation $W(t)$, which is relevant in the case where $W(t)$ has 
infinitely many frequencies which may resonate with the continuous spectrum.
In section \ref{se:application} we apply the results of section \ref{se:mainresults} to the case of Schr\"odinger 
operators $H_0\ =\ -\Delta + V(x)$ defined on $L^2(\R^3)$. To check the key 
local energy decay hypotheses we use results of Jensen and Kato 
 \cite{kn:JK}  on expansions of the resolvent of $H_0$ near zero energy, 
 the edge  of the continuous spectrum. In section \ref{se:normalform} the dynamical system 
 (\ref{eq:perturbed}) is reformulated as a system governing the interaction
of the bound state, and dispersive part  of the solution.  This 
 section contains an important 
 computation, in which the key resonance is made explicit and a  perturbed 
"normal form" for the bound state evolution is derived (Proposition 4.1).
 Sections \ref{se:ode} and \ref{se:localdecay} contain estimates for the bound state and dispersive 
parts of the solution for intermediate and large time scales. In section 
\ref{se:generalizations} we discuss extensions of our 
Theorems \ref{th:twofive}-\ref{th:twotwo} to a more general class of 
perturbations. We shall 
frequently make use of some singular operators which are rigorously defined in section \ref{se:sop}, an appendix, and of elements of the theory of almost periodic functions
 \cite{kn:Bohr, kn:Levitan}, 
 which are assembled in section \ref{se:app}, the second appendix. 
\bigskip

\nit {\bf Notations and terminology:}

\nit Throughout this paper we will use the following notations:

\nit $\N\ =\{1,\ 2,\ 3, \ldots\};$ 

\nit $\N_0\ =\{0,\ 1,\ 2,\ 3,\ldots\};$

\nit $\Z\ =\{\ldots , -3,\ -2,\ -1,\ 0,\ 1,\ 2,\ 3,\ldots\};$

\nit for $z$ a complex number, $\Re z$ and $\Im z$ denote, respectively, its
real and imaginary parts;

\nit a generic constant will be denoted by $C$, $D$, etc;

\nit $\langle x\rangle\ =\ \left(
1+|x|^2 \right)^{1\over2}$;

\nit ${\cal L}({\cal A,B}) =\ $ the space of bounded linear operators from ${\cal A}$ to ${\cal B}$; ${\cal L(A,A)}\equiv {\cal L(A)} $.

\nit Functions of self-adjoint operators are defined via the spectral theorem;
see for example \cite{kn:RS1}. The operators containing boundary value of resolvents or singular distributions applied to self-adjoint operators are defined in section \ref{se:sop}.

\bigskip\noindent {\bf Acknowledgements:} This research was supported in part
by National Science Foundation grant DMS-9500997. Part of this work was done while E. Kirr participated in the Bell Labs/Lucent Student Intern Program. The authors wish to thank A. Soffer and P.D. Miller for discussions on this work.

\section{General formulation and main results.}\label{se:mainresults}

Consider the general system
\ba
i \D_t\phi(t) & = & \left(H_0 + W(t)\right)\phi(t) , \no \\
\phi|_{t=0} & = & \phi(0). \label{eq:generalse}
\ea
Here, $\phi(t)$ denotes a function of time, $t$, with 
values in a complex Hilbert space ${\cal H}$. 

\nit\underline{\bf Hypotheses on $H_0$:}

\nit{\bf (H1)} $H_0$ is self-adjoint 
on ${\cal H}$ and both $H_0$ and $W(t),\ t\in\R^1,$ are 
 densely defined on a subspace ${\cal D}$ of ${\cal H}$. 
 
\nit The norm on
 ${\cal H}$ is denoted by $\|\cdot\|$, and the inner product of $f,g\in
 {\cal H}$, by $\left(f,g\right)$.

\nit{\bf (H2)} 
The spectrum of  $H_0$  is assumed to consist of an absolutely continuous part,
 $\sigma_{\rm cont}(H_0)$, with associated spectral projection 
 $\Pc$ and a  single isolated eigenvalue $\lambda_0$ 
 with corresponding normalized eigenstate, $\psi_0$, {\it i.e.}
\be H_0\psi_0\ =\ \lambda_0 \psi_0,\ \| \psi_0\|=1.\ee

The manner in which we shall measure the decay of solutions is 
 typically in a local decay sense, {\it e.g.} 
 for the scalar Schr\"odinger equation governing a function defined on $\R^n$ we measure local decay using
 the norms:
$f\mapsto \| \langle x\rangle^{-s} f \|_{L^2}$, where $s>0$. 
 So that our theory applies  to a class of general systems (involving,
 for example, 
 vector equations with matrix operators), we assume the existence of 
 self-adjoint "weights", $\wmi$ and $\wpl$ such that 

\ (i)\ $w_+$ is defined on a dense subspace of ${\cal H}$ and on 
 which $w_+\ge cI$,
 \ $c > 0$.

\ (ii)\  $w_- \in {\cal L}({\cal H})$ such that $Range(\wmi)\subseteq Domain(\wpl)$.

\ (iii)\ $\wpl\ \wmi\ \Pc\ =\ \Pc\ $on ${\cal H}$ and $\Pc\ =\ \Pc\ \wmi\ \wpl$ on the domain of $\wpl$.

In the scalar case $\wpl$ and $\wmi$ 
 correspond to multiplication by 
  $ \langle x\rangle^s$ and    $\langle x\rangle^{-s}$, 
 respectively, see section \ref{se:application}.

The following hypothesis ensures that the unperturbed dynamics
satisfies sufficiently strong dispersive time-decay estimates. Let 
$\left\{\mu_j\right\}_{j\in\Z}$ denote the set of Fourier exponents associated
with the perturbation $W$ (see hypothesis {\bf (H4)} below).

\nit {\bf (H3)} \underline{Local decay estimates on $e^{-iH_0t}$}:
Let $r_1>1$.
 There exist $w_+$ and $w_-$, as above, and a constant $\cal C$ such that 
for all $f\in {\cal H}$ satisfying $\wpl f\in {\cal H}$ we have:

\nit 
\ba
{\rm {\bf (a)}}\ \ \|  \wmi e^{-iH_0t} {\bf P_c} f \| & \le & 
{\cal C}\ \lan t\ran^{-r_1} \|\wpl f\|,\ {\rm for}\ t\in\R;
\label{eq:localdecay}\\
{\rm {\bf (b)}}\ \ \|\wmi e^{-iH_0t}(H_0-\lambda_0 - \mu_j -i0)^{-1}\
 \Pc f\| & \le &
 \ {\cal C}\ \lan t\ran^{-r_1}\ \|\wpl f\|,\ {\rm for}\ t\ge 0
  \label{eq:singularldest}
\ea
and for all $j\in\Z$. For $t<0$ estimate (\ref{eq:singularldest})
is assumed to hold with $-i0$ replaced by $+i0$. See section \ref{se:sop} for the definition of the singular operator in (\ref{eq:singularldest})

\begin{rmk}\label{rmk1} 
 There is a good deal of literature on local energy decay estimates of the form
 form (\ref{eq:localdecay}) for 
$e^{-iH_0t}\Pc$ in the case $H_0=\ -\Delta+V(x)$ on $L^2(\R^n)$. These results
require sufficient regularity and decay of the potential $V(x)$. We refer the
reader to \cite{kn:JK}, \cite{kn:JSS} and \cite{kn:Murata}; see also
\cite{kn:Rauch}, \cite{kn:Schonbek}.
\end{rmk}
\medskip


\begin{rmk} \label{rmk2}
Estimates of the type {\bf (H3b)} are obtained in \cite[Appendix A]{kn:nh, kn:TDRT1}. A key point here is that we require that one can choose the constant, 
 ${\cal C}$, in
 (\ref{eq:singularldest}) to hold for all $\mu_j$. It appears difficult to deduce
this uniformity of the constant by the general arguments used in \cite{kn:nh}
and \cite{kn:TDRT1}. However, in section \ref{se:application}, where we apply our results to a 
 class of Schr\"odinger operators, we can verify ${\bf (H3b)}$ using known results on the spectral measure.
\end{rmk}
\medskip

\nit {\bf (H4)} \underline{\bf Hypotheses on the perturbation $W(t)$:}

\nit We consider  
 time-dependent symmetric perturbations of the form
\be
W(t) = {1\over2}\beta_0+\sum_{j\in\N}\cos (\mu_j t)\  \beta_j,\  {\rm with}\   
\beta_j^*=\beta_j\  {\rm and}\  \sum_{j\in\N_0}\|\beta_j\|_{{\cal L}({\cal H})}<\infty.
\label{eq:generalW}
\ee

In many applications, $\beta_j$ are spatially
localized scalar or matrix function. Note that formula
(\ref{eq:generalW}) can be rewritten in the form:
\be
W(t) = {1\over2}\  \sum_{j\in\Z}\exp (-i\mu_j t)\beta_j,
\label{eq:generalW1}
\ee
where, $\mu_0=0$ and for $j<0$, $\mu_j=-\mu_{-j},\ \beta_j=\beta_{-j}$. 
 Thus, $W(t)$ is
an almost periodic function with values in the Banach space 
 ${\cal L}({\cal H})$, with
the Fourier exponents $\left\{\mu_j\right\}_{j\in\Z}$ and corresponding
  Fourier
coefficients $\left\{\beta_j\right\}_{j\in\Z}$; 
 see, for example, \cite{kn:Levitan}. 

To measure the size of the perturbation  $W$, we introduce the norm
\be
||| W |||\ \equiv\ {1\over 2}\sum_{j\in\Z}\| \wpl\; \beta_j\|_{{\cal L}({\cal H})}
\ +\ {1\over 2}\sum_{j\in\Z}\|\ \beta_j\ \|_{{\cal L}({\cal H_-,H_+})},
\label{eq:betanorm}
\ee
which is assumed to be finite. Here ${\cal H_+}$, respectively ${\cal H_-}$, denote the closure of the domain of $\wpl$, respectively the range of $\Pc $, with norm $f\rightarrow\|\wpl f\|$, respectively $f\rightarrow\|\wmi f\|$.

\begin{rmk}\label{gbeta}
\noindent A special case which arises in various models, is:
\be 
W(t)\ =\ g(t)\beta,\label{eq:specialcase}
\ee
where 
\be g(t)\ =\ \sum_{j}\ g_j\cos{\mu_j t},\no\ee
 $\|w_+\beta\|_{\cal L(H)} +\ \|\beta \|_{\cal L(H_-,H_+)}\ <\ \infty$ and the sequence
$\{g_j\}$ is absolutely summable. 
\end{rmk}
\begin{rmk} Our results are valid in the more general case
$$ W(t) =  {1\over2}\beta_0+\sum_{j\in\N}\cos (\mu_j t + \delta_j) \  \beta_j,$$
where $\beta_j$ are self-adjoint such that expression (\ref{eq:betanorm}) is
finite. This follows because
  the proofs use only the self-adjointness of $W$ and the
expansion:
$$
W(t) = {1\over2}\  \sum_{j\in\Z}\exp (-i\mu_j t)\tilde\beta_j, 
$$
where $\tilde\beta_j=e^{-i{\rm sgn}(j)\delta_j}\beta_j$ and 
 $\mu_{-j}=-\mu_j,\ \mu_0=0$.
\end{rmk}

We will impose a resonance condition which says that 
 $\left\{\la_0+\mu_j\right\}_{j\in\Z}\cap\sigma_{\rm cont}(H_0)$ is nonempty 
 and that there is nontrivial coupling; see section 1.4. Let us first denote by $\res$ the 
following set:
\be \res\ =\ \{j\in\Z\ :\ \la_0+\mu_j\in\spectrum\}. \label{eq:defres}\ee
 
\nit {\bf (H5)} \underline{Resonance condition - Fermi golden rule}
$\res$ is nonempty and furthermore, there exists $\theta_0>0$, independent of $W$,  such that 
\be
\Gamma\ \equiv\ {\pi\over4}\ \sum_{j\in\res}
  \left( \Pc\beta_j\psi_0, \; \delta(H_0 - \la_0 - \mu_j)\Pc \beta_j\psi_0
\right) \ge
 \theta_0||| W |||^2 > 0
 \label{eq:tdfgr}
 \ee

\begin{rmk}\label{rmk3} For the exact definition of the Dirac type operator in (\ref{eq:tdfgr}) see section \ref{se:sop}.
That $\Gamma$ is finite is a consequence 
 of the estimate (\ref{eq:deltabound}) and
\be \label{eq:gammaeval}
\Gamma\le {C_0\over\pi}\sum_j\|\wpl\beta_j\|^2\le {C_0\over\pi} |||W|||^2
\ee
\end{rmk}
\medskip

  

We now state our main results:
\begin{theo}\label{th:twofive}
 Let us fix $H_0$ and $W(t)$ satisfying hypotheses {\bf (H1)-(H5)}. Consider the initial value problem:
\ba
i \D_t\phi(t) & = & \left(H_0 +\ve W(t)\right)\phi(t) , \no \\
\phi|_{t=0} & = & \phi(0), \label{eq:epsgeneralse}
\ea
with $w_+\phi(0)\in{\cal H}$. 
Then, there exists an $\ve_0 >0$ (depending on ${\cal C},\ r_1,$ and $\theta_0$)
such that whenever $|\ve |< \ve_0$, 
 the solution, $\phi(t)$, of (\ref{eq:epsgeneralse}) satisfies the local decay
estimate:
\begin{equation}
\| \wmi\ \phi(t)\| \le C\lan t\ran^{-r_1} \| \wpl\ \phi(0)\|,\ \ t\in\R.
\label{eq:ldestimate}
\end{equation}
\end{theo}

Under the same hypotheses as Theorem \ref{th:twofive}, we obtain   
 more detailed information on the behavior of $\phi(t)$: 
 
\begin{theo}\label{th:twosix} 
Assume the hypotheses of Theorem \ref{th:twofive}.
  For any $0<\gamma <\Gamma$ there exist the constants $C$ and $D$ 
(depending
on ${\cal C},\ r_1,\ \theta_0$  and $\gamma$) such that any solution of 
(\ref{eq:epsgeneralse}), for $|\ve|<\ve_0$ and $w_+\phi(0)\in{\cal H}$, satisfies:
\begin{eqnarray}
\phi(x,t) &=& a(t)\psi_0 + \phi_d(t),\ \ \left(\psi_0\ ,\ \phi_d(t)\right)\ 
   =\ 0,\nn\\
a(t) &=&\  
a(0)\ e^{-\ve^2 (\Gamma-\gamma )|t|}e^{i\omega(t)}\ 
 +\ R_a(t)\nn \\
P(t) &=&\ 
P(0)\ e^{-2\ve^2 (\Gamma-\gamma )|t|}\ +\ R'_a(t)\nn \\
\phi_d(t) &=& e^{-iH_0t}\ \Pc \phi(0)\ +\ {\tilde \phi(t)}.\label{eq:projection1}
\end{eqnarray}
where $\Gamma$ is given by (\ref{eq:tdfgr}) and $\omega(t)$ is a real-valued phase given by
\ba
\omega(t)\ &=&\ \la_0t
 -\ve\left(\psi_0,\ \int_0^t W(s)ds\ \psi_0\right)
\no\\
&+& {1\over4}\ve^2 t\sum_{j\in\Z} \left(\beta_j\psi_0,{\rm P.V.}
  (H_0-\lambda_0-\mu_j)^{-1}\Pc\beta_j\psi_0\right), \no\\
&+&\
 {1\over4}\ve^2\ \Re\int_0^t\sum_{j,k\in\Z,j\neq k}e^{i(\mu_k-\mu_j)t}\left(\beta_k\psi_0,
  (H_0-\lambda_0-\mu_j-i0)^{-1}\Pc\beta_j\psi_0\right).\label{eq:omegadef}
\ea
$P(t)$ is defined in (\ref{eq:Pdef})
 and for any fixed $T_0>0$ we have 
\ba
\left| R_a(t)\right|\ &\le &\ C\ |\ve|\ |||W|||,\ |t|\le \frac{T_0}{\ve^2 \Gamma}
 \label{eq:Raestimate}\\
\left| R'_a(t)\right|\ &\le &\ D\ |\ve|\ |||W|||,\ |t|\le \frac{T_0}{\ve^2 \Gamma}
 \label{eq:Raprimestimate}
\ea
 Moreover, 
$$
 |R_a(t)|={\cal O}(\lan t\ran^{-r_1}),\ 
 |R'_a(t)|={\cal O}(\lan t\ran^{-r_1}),\ |t|\to\infty .
$$
Finally, ${\tilde \phi}= \phi_1 + \phi_2$ is  given in 
 (\ref{eq:intphideqn}), with 
 $\| w_-\tilde\phi(t)\|\ =\ {\cal O}(\lan t\ran^{-r_1})$ as $|t|\to\infty$.
Therefore, by {\bf (H3)} same holds  
$\| w_-\phi_d(t)\|\ =\ {\cal O}(\lan t\ran^{-r_1})$ as $|t|\to\infty$.
\end{theo}
\medskip

\begin{rmk}\label{rmk12}
 Suppose the initial data is given by the bound state of the
unperturbed problem, {\it i.e.} $\phi(x,0)=\psi_0(x)$, $a(0)=1$,\ $\phi_d(0)=0$.
 Then, from the
expansion of the solution 
 we have that for  $0\le t\le \ve^{-2}\Gamma^{-1}$ that $P(t)$ (see (\ref{eq:Pdef}))
is of order $e^{-2\ve^2(\Gamma-\gamma) t}$, with an error of order $\ve $.
Hence it is natural to view the state $\psi_0e^{-i\lambda_0 t}$ as a
{\rm metastable state}, with lifetime $\tau =\ve^{-2}(\Gamma-\gamma)^{-1}\sim
 \ve^{-2} |||W|||^{-2}$. Although $\gamma>0$ is arbitrary
 we have not inferred that the actual lifetime is $\tau=\ve^{-2}\Gamma^{-1}$ under 
hypothesis {\bf (H1)-(H5)}.
 The reason is that the constants $C$ and $D$ in the estimates 
(\ref{eq:Raestimate}) and (\ref{eq:Raprimestimate})
 blow up as $\gamma\searrow 0$. In order to remedy this we need an additional 
 hypothesis:
\end{rmk}

\nit {\bf (H6)} \underline{Control of small denominators}: 
 There exists $\xi >0$, independent of $W$, such that  
\be
\sum_{j\in\res,\ k\in\Z,\ j\neq k}
\left|\frac{1}{\mu_j-\mu_k} 
(\Pc\beta_k\psi_0,\delta(H_0-\la_0-\mu_j)\Pc\beta_j\psi_0)\right|\le
\xi\ |||W|||^2.\label{eq:summability}
\ee

\begin{rmk}\label{rmk4}
By (\ref{eq:deltabound}) we have
\be \sum_{j\in\res,\ k\in\Z,\ j\neq k}\left|
(\beta_k\psi_0,\delta(H_0-\la_0-\mu_j)\beta_j\psi_0)\right|
 \le C\ \pi^{-1}\ |||W|||^2 \label{eq:deltaeval}\ee
is finite (see also Remark \ref{rmk3}). Thus, {\bf (H6)}  
is important only if:
\be
\inf\{|\mu_j-\mu_k|:\ j,k\in\Z,\ j\neq k\ {\rm and}\
\la_0+\mu_j\in\spectrum\}=0,\label{eq:accumulation}
\ee
{\it i.e.} the Fourier exponents $\{\mu_j\}$ are such that $\lambda_0+\mu_j$ 
accumulate  in $\sigma_c$.
 In particular, if the perturbation $W(t)$ consists of a trigonometric polynomial:
\be W(t)\ =\ \sum_{j=1}^N\ \cos{\mu_j t}\ \beta_j,\no\ee
then {\bf  (H6)} is trivially satisfied.
\end{rmk}

\begin{rmk}\label{rmk6}
Hypothesis  {\bf (H6)} can be imposed by balancing the clustering of the 
frequencies $\lambda_0+\mu_j$ in the continuous spectrum of 
 $H_0$ with rapid decay of $\left(\beta_k\psi_0\ ,\ \delta(H_0-\lambda_0-\mu_j)\
 \beta_j\psi_0\right)$ as $j,k\to\infty$. Let $\beta_j(x) = 
  g_j\ \beta(x)$. Then, $W(t,x)=\sum_j\ g_j\cos(\mu_j t)\ \beta(x)$. Using 
Remark \ref{rmk3} 
we find that the left hand side of  (\ref{eq:summability}) is bounded by 
 $\sum_{j,k\in\Z; j\ne k} |g_j g_k|\ |\mu_j-\mu_k|^{-1}\ |||W|||^2$. The
constant $\xi$ in  (\ref{eq:summability})  is finite if, for example,  
 $\mu_j = 2|\lambda_0|+|j|^{-1},\ g_j =  |j|^{-2-\tau},\ \tau>0$.
\end{rmk}

In case {\bf (H6)} is satisfied we have the following improvement of 
Theorem \ref{th:twosix}:
\begin{theo}\label{th:twotwo} 
  Assume the hypotheses {\bf (H1)-(H6)} hold. Then
there exist $\ve_0$ and the constants $C,\ D$ (depending
on ${\cal C},\ r_1,\ \theta_0$ and $\xi$) such that any solution of 
(\ref{eq:epsgeneralse}), for $|\ve |<\ve_0$ and $w_+\phi(0)\in{\cal H}$, satisfies
\begin{eqnarray}
\phi(x,t) &=& a(t)\psi_0 + \phi_d(t),\ \ \left(\psi_0\ ,\ \phi_d(t)\right)\ 
   =\ 0,\nn\\
a(t) &=&  a(0)\ e^{-\ve^2\Gamma |t|}\ 
 e^{i\omega(t)}\  +\ R_a(t)\nn \\
P(t) &=&  P(0)\ e^{-2\ve^2\Gamma |t|}\ +\ R'_a(t)\nn \\
\phi_d(t) &=& e^{-iH_0t}\ \Pc \phi(0)\ +\ {\tilde \phi(t)}.\label{eq:projection}
\end{eqnarray}
Here, $\omega(t)$ is given by (\ref{eq:omegadef}) and 
$R_a(t),\ R'_a(t),\ \wmi\phi_d(t)$ satisfy the estimates of Theorem \ref{th:twosix} .
\end{theo}

%

\section{ An application: the Schr\"odinger equation }\label{se:application}

In this section we verify hypotheses {\bf (H1)-(H4)} in 
the particular case of the Schr\"odinger equation on the three dimensional space 
with a time almost periodic and spatially localized perturbing potential:
\be
i\D_t\phi\ =\ \left(-\Delta+V(x)\right)\phi\ +\ \ve W(x,t)\phi,\label{eq:appgen}
\ee
with
$
\phi:\R^3\times\R\rightarrow\C,\ 
(x,t)\rightarrow\phi(x,t)
$
and $$W(x,t)={1\over2}\beta_0(x)+\sum_{j\in\N}\cos(\mu_jt)\beta_j(x)$$ where $\mu_j\in\R,\ j\in\N_0,$ and $\beta_j:\R^3\rightarrow\R,\ j\in\N,$ are localized functions.
Models of the sort considered in this example occur in the study of
ionization
of an atom by a time-varying electric field; see \cite{kn:L-L},
\cite{kn:G-P}.

 We take ${\cal H}=L^2(\R^3)$, and 
 $H_0\ \equiv\ -\Delta\ +\ V(x)$, where $V(x)$ is real-valued with moderately
 short range. More precisely, we suppose that there exists $\sigma>4$ and a
 constant $D$ such that
 \be
 |V(x)|\leq D(1+|x|)^{-\sigma}.\label{eq:potentialdecay}
 \ee
 Thus, $H_0$ is self-adjoint and densely defined in
 $L^2$.
In what follows we assume that $H_0$ has exactly one eigenvalue which is strictly
negative and that the remainder of the spectrum is  absolutely 
continuous and equal to the positive half-line. 
 Our results can be extended to operators with strictly negative, multiple eigenvalues \cite{kn:KW}.

We first discuss the local decay hypothesis  {\bf (H3)}. As weights used to
measure local energy decay we take 
$w_\pm \equiv \lan x\ran^{\pm s}$ where $s>7/2$ and fix $r_1=3/2$. Our aim
is to obtain the estimates:
\ba
{\rm {\bf (H3 a)}}\ \ \|  \wmi e^{-iH_0t} {\bf P_c} f \| & \le & 
{\cal C}\ \lan t\ran^{-3/2} \|\wpl f\|,\label{eq:H3a}\\
{\rm {\bf (H3 b)}}\ \ \|\wmi e^{-iH_0t}(H_0-\lambda_0 - \mu_j -i0)^{-1}\
 \Pc f\| & \le &
 \ {\cal C}\ \lan t\ran^{-3/2}\ \|\wpl f\|,\label{eq:H3b}\
\ea
for all $\mu_j\in\Z$, with ${\cal C}$ independent of $j$.

We shall assume that the frequencies $\{ \lambda_0+\mu_j \}$ do not accumulate 
at zero, the edge of the continuous spectrum of $H_0$:
\be
m_*\ \equiv\  \min\{\ |\lambda_0+\mu_j|\ :\ j\in\Z\ \} >\ 0.\no\ee

To prove (\ref{eq:H3a}) and (\ref{eq:H3b}) we use
the spectral representation for the operators $e^{-iH_0t}\ \Pc$ and 
$e^{-iH_0t}\ (H_0-\la_0-\mu_j-i0)^{-1} \Pc$, namely:
\ba 
e^{-iH_0t}\Pc &=&\int_0^\infty e^{-i\la
t}E'(\la )d\la \label{eq:srwave}\\
e^{-iH_0t}(H_0-\la_0-\mu_j-i0)^{-1}\Pc &=&\int_0^\infty e^{-i\la
t}(\la-\la_0-\mu_j-i0)^{-1}E'(\la )d\la \label{eq:srsingularwave}
\ea
where $E'(\la )=\pi^{-1}\ \Im (H_0-\la-i0)^{-1}$ is the spectral 
 density induced by 
 $H_0$, \cite{kn:JK}. 

The technique of getting {\bf (H3a)} from (\ref{eq:srwave}) is presented in
\cite[Section 10]{kn:JK} and it can be summarized in the following way.
We decompose the integral in (\ref{eq:srwave}) in two parts, corresponding
to low energies ($\lambda$ near zero) and high energies 
 ($\lambda$ away from zero) by writing
\ba
E'&=&\chi E'+(1-\chi)E'\no\\
\wmi e^{-iH_0t}\Pc\wmi &=&\int_0^\infty e^{-i\la
t}\chi(\la) \wmi E'(\la )\wmi d\la + \int_0^\infty e^{-i\la
t}(1-\chi (\la)) \wmi E'(\la )\wmi d\la\no\\
&=&\ S_1\ +\ S_2.\label{eq:dsrwave} 
\ea
Here $\chi(\lambda)$ is a smoothed characteristic function of 
 a neighborhood of origin, chosen so that 
\ba
\chi(\lambda)\ &\equiv&\ 1,\ |\lambda|\ \le\ {1\over2}m_*\no\\
\chi(\lambda)\ &\equiv&\ 0,\ |\lambda|\ \ge\ {3\over4}m_*.\no
\ea

To estimate the two integrals in (\ref{eq:dsrwave}) we make use of the detailed
results of \cite{kn:JK} on the family of operators $\{\ E'(\lambda)\ \}$. 
First, by Theorem 8.1 and Corollary 8.2 of \cite{kn:JK}, 
 $\wmi \D_\lambda^kE'(\la)\wmi$ is bounded on $L^2$ and satisfies
\be
 \|\ \wmi \D_\lambda^kE'(\la)\wmi\ \|_{{\cal B}(L^2)}
  = O(\la^{-(k+1)/2})\ {\rm as}\ \la\rightarrow\infty,
\label{eq:E3est}\ee
for $k\in\{0,1,2,3\}$.
Integration by parts twice in
  the second integral in (\ref{eq:dsrwave})  
 and use of the estimate (\ref{eq:E3est}) with $k=2$ 
yields the estimate:
\be \|\ S_2\ \|_{{\cal B}(L^2)}\ =\ 
 o(t^{-2})\ {\rm as}\ t\rightarrow\infty.\no\ee

Next, by Theorem 6.3 of \cite{kn:JK} we have the low energy 
asymptotic expansion:
\be
\wmi E'(\la)\wmi =-\la^{-1/2}B_{-1}+\la^{1/2}B_1+o(\la^{1/2})\ {\rm as}\
\la\rightarrow 0,\label{eq:lowenergy}
\ee
where $B_{-1},\ B_1$ are bounded linear operators on $L^2$. 
Use of this expansion in the first integral of (\ref{eq:dsrwave}) yields
the expansion in ${\cal B}(L^2)$:
\be
 S_1\ =\ (\pi i)^{-1/2}t^{-1/2}B_{-1}-(4\pi i)^{-1/2}t^{-3/2}B_1+o(t^{-3/2})\ {\rm as}\
t\rightarrow\infty. \label{eq:tasymptotic}
\ee

Thus {\bf (H3a)} is satisfied provided that $B_{-1}$ is the null operator or
equivalently $H_0\psi=0$ has no solution with the property $\wmi\psi\in
L^2(\R^3)$. The last condition holds for generic potentials $V(x)$ and when
it is violated one says that $H_0$ has {\it zero energy resonance}; see
\cite{kn:JK} for details.

In the same way one can prove {\bf (H3b)} from the spectral representation
(\ref{eq:srsingularwave}) provided that the integral is non-singular, i.e.
$\la_0+\mu_j<0$. In the case $\la_0+\mu_j \ge m_*> 0$  we first decompose the
singular integral in two parts, one away from singularity point, $\la_0+\mu_j$,
and the other in a neighborhood of it by using the smoothed characteristic
function 
\be \chi_j(\lambda)\ =\ \chi(\lambda-\lambda_0-\mu_j),\no\ee
 which is supported in a neighborhood of $\la_0+\mu_j$, which does not include
$\lambda=0$:
\ba
e^{-iH_0t}(H_0-\la_0-\mu_j-i0)^{-1}\Pc &=&\int_0^\infty e^{-i\la
t}(\la-\la_0-\mu_j)^{-1}(1-\chi_j(\la))E'(\la )d\la \no\\
&+& \int_0^\infty e^{-i\la
t}(\la-\la_0-\mu_j-i0)^{-1}\chi_j(\la)E'(\la )d\la,
\label{eq:dsrsingularwave}
\ea
The non-singular integral may be treated as above while the singular one
defines the  singular operator:
$$ 
T_j = e^{-iH_0t}(H_0-\la_0-\mu_j-i0)^{-1}\chi_j(H_0)\Pc,
$$
via the spectral theorem. Here, $T_j=\lim_{\eta\searrow 0}T_j^{\eta}$, where
$$T_j^\eta\equiv e^{-iH_0t}(H_0-\la_0-\mu_j-i\eta)^{-1}\chi_j(H_0)\Pc.$$ 
To estimate its $L^2$ operator norm we use the  integral representation
\be
\wmi T_j^\eta \wmi = \frac{1}{i}\int_t^\infty e^{i(\la_0+\mu_j+i\eta)(s-t)}\wmi e^{-iH_0s}
\chi_j(H_0)\Pc\wmi ds. \label{eq:hufhuf}
\ee
But this reduces to the evaluation of 
\be
\wmi e^{-iH_0s}
\chi_j(H_0)\Pc\wmi =\int_0^\infty e^{-i\la s}
\chi_j(\la)\wmi E'(\la)\wmi d\la,\ s\ge t,\label{eq:huf1}
\ee
where we used again the spectral representation theorem. 
Integration by parts three times in (\ref{eq:huf1}) and use of the estimate (\ref{eq:E3est})
with $k=3$ implies
$$
\|\wmi e^{-iH_0s}
\chi_j(H_0)\Pc\wmi\|_{{\cal B}(L^2)}=o(s^{-3})\ {\rm as}\ t\rightarrow\infty.
$$
 Replacing this in
(\ref{eq:hufhuf}), integrating and passing to the limit as $\eta\searrow 0$ we
obtain an $o(t^{-2})$ estimate for $T_j$ which is even better than we need
to satisfy {\bf (H3b)}.

Moving now towards hypothesis {\bf (H4)}, we may choose the time-dependent perturbation to be of the form:
\be 
W(x,t) \ ={1\over2}\beta_0\ +\ \sum_{j\in\N}\cos{\mu_jt}\ \beta_j(x),
\no\ee
with $\beta_j$  rapidly decaying in $x$, {\it e.g.}
 $\lan x\ran^{2s}\|\beta_j(x)\|\le C_j$ for all $x\in\R^3,\ j\in\N_0$, where
 $\sum_{j\in\N_0} C_j<\infty$. Thus, {\bf (H4)} is satisfied as well. 
 
Therefore, our main results  
Theorems \ref{th:twofive}-\ref{th:twosix} on the  structural instability of the
unperturbed bound state, and large time behavior for systems of the form (\ref{eq:appgen}) apply provided {\bf (H5)}, the Fermi Golden Rule resonance condition, holds. For results concerning more general perturbations than the ones in (\ref{eq:appgen}) see section \ref{se:generalizations}.
\section{ Decomposition and derivation of the dispersive normal form}
\label{se:normalform}

The results of this section rely on hypothesis {\bf (H1)} through {\bf (H4)}
only, so they may and will be used in proving  
Theorems \ref{th:twofive}-\ref{th:twotwo}.

 As in \cite{kn:TDRT0}, \cite{kn:nh} and \cite{kn:TDRT1}, we begin by deriving 
 a decomposition of the solution, $\phi(t)$, 
which will facilitate the study of its large time behavior.
 Let  
 \begin{equation}
\phi(t) = a(t) \psi_0 + \phi_d(t), \label{eq:ansatz}
\end{equation}
with the orthogonality condition
\begin{equation}
(\psi_0, \phi_d(t)) = 0 \ \ {\rm for \ all} \ \ t \label{eq:orthog}.
\end{equation}
Note therefore that $\phi_d\ =\ \Pc\phi_d$.
\medskip

We proceed by first inserting (\ref{eq:ansatz}) into (\ref{eq:epsgeneralse}),
which 
yields the equation:
\ba
i\D_ta(t)\psi_0\ +\ i\D_t\phi_d(t)\ &=&\ \lambda_0a(t)\psi_0\ + H_0\phi_d(t)
\no\\
&+&\ \ve a(t)W(t)\psi_0\ +\ \ve W(t)\phi_d(t)\label{eq:substit}
\ea

Taking the inner product of (\ref{eq:substit}) with  $\psi_0$ we get the 
  following equation for $a(t)$:
\ba
i \D_t a\ &=&\ \lambda_0 a(t)\ +\  \ve\left(\psi_0, W(t)\psi_0\right) a(t)\ +\
\ve \left(\psi_0, W(t)\phi_d\right),\label{eq:aeqn1}\\
 a(0)\ &=&\ \left(\psi_0,\phi(0)\right)\no
\ea
In deriving (\ref{eq:aeqn1}) we have used that $\psi_0$ is normalized and
the relation
\be \left(\psi_0 , \D_t\phi_d\right) = 0,\nonumber\ee
a consequence of (\ref{eq:orthog}).

Applying $\Pc$ to (\ref{eq:substit}), we obtain an equation for
$\phi_d$:
\ba
i \D_t \phi_d(t)\ &=&\ H_0 \phi_d(t)\ +\ \ve\Pc W(t)\phi_d(t)\ +\ \ve a(t)\Pc
W(t)\psi_0,\label{eq:phideqn}\\
\phi_d(0)\ &=&\ \Pc\phi(0)\no
\ea
Since we are after a slow resonant decay phenomenon, it will prove
advantageous to extract the fast oscillatory behavior of $a(t)$. 
We therefore define:
\be    A(t) \equiv e^{i\la_0 t} a(t).\label{eq:Adef}\ee
Then, (\ref{eq:aeqn1}) reads
\begin{equation}
\D_tA\ =\  -i\ve A\left(\psi_0, W(t)\psi_0\right)\ -\ i\ve e^{i\la_0 t}\ 
 \left(\psi_0, W(t)\phi_d(t)\right).\label{eq:Aeqn1}
\end{equation}
Solving (\ref{eq:phideqn}) by Duhamel's formula we have
\ba
\phi_d(t) &=& e^{-iH_0t}\phi_d(0)\ -\ 
 i \ve\int^t_0 e^{-iH_0(t-s)}  \Pc W(s) a(s) \psi_0 ds \nonumber \\
&-&   i \ve\int^t_0 e^{-iH_0(t-s)} \Pc  W(s) \phi_d(s)\ ds\no \\
&\equiv& \phi_0(t)\  +\ \phi_1(t)\  +\ \phi_2(t) \label{eq:intphideqn} .
\ea
By standard methods, the system 
 (\ref{eq:Aeqn1})-(\ref{eq:intphideqn}) for $A(t)$
and $\phi_d(t) = \phi(t) - e^{-i\lambda_0t}\ A(t)\ \psi_0$
has a global solution in $t$ with 
$$A\in C^1(\R),\  
 \|\phi_d(t)\|\in C^0(\R),\  \|w_-\phi_d(t)\|\in C^0(\R).
$$
 Our analysis of the $|t|\to\infty$ behaviour is based on a study of
 this system.

By inserting (\ref{eq:intphideqn}) into (\ref{eq:Aeqn1}) we get
\begin{equation}
\D_tA(t)\ =\ -i\ve A(t) \left(\psi_0, W(t) \psi_0\right)\ -\ 
 i\ve e^{i\la_0t} \sum^2_{j=0} \left(\psi_0, W(t)\phi_j\right).\label{eq:Aeqn2}
\end{equation}

We next give a detailed expansion of the sum in
(\ref{eq:Aeqn2}). It is in the  $j=1$ term
  that the key
resonance is found. This makes it  possible to find a normal form for
(\ref{eq:Aeqn2}) in which {\it internal damping} in the system 
 is made explicit. This damping reflects the transfer of energy
 from the discrete to continuum modes of the system and the associated
 radiative decay of solutions.

\begin{prop}\label{pr:Aequation} For $t>0$,
\be
\D_t A(t)\ =\ \left(\ -\ve^2\Gamma\ +\  \rho(t)\ \right)\ A(t)\ +\ E(t), 
\label{eq:Aeqn3}
\ee
where $\Gamma$ is defined in (\ref{eq:tdfgr}),
\ba \rho(t)\ &=&\ 
 -i\ve\left(\psi_0,\ W(t)\psi_0\right)\ +
\no\\
  &+&\ {i\over4}\ve^2\sum_{j\in\Z} \left(\beta_j\psi_0,{\rm P.V.}
  (H_0-\lambda_0-\mu_j)^{-1}\Pc\beta_j\psi_0\right)\no\\
&+&\ 
{i\over4}\ve^2\sum_{j,k\in\Z,j\neq k}e^{i(\mu_k-\mu_j)t}\left(\beta_k\psi_0,
  (H_0-\lambda_0-\mu_j-i0)^{-1}\Pc\beta_j\psi_0\right)\label{eq:rhoexpansion}
\ea
and 
\ba
\lefteqn{E(t)\ =\  
  -{i\over4}\ve^2 A(0)e^{i\lambda_0 t}\sum_{j,k\in\Z}e^{i\mu_k t}\ 
 \left(\beta_k\psi_0,\ e^{-iH_0t}\ (H_0 - \la_0 -\mu_j-i0)^{-1} 
 \Pc \beta_j\psi_0\right)}\no  \\ 
  & & -{i\over 4}\ve^2 e^{i\la_0 t}\sum_{j,k\in\Z}e^{i\mu_k t}
 \left(\beta_k\psi_0, \int_0^t e^{-iH_0(t-s)}(H_0 - \la_0 -\mu_j-i0)^{-1}
\Pc e^{-i(\la_0+\mu_j)s}\D_sA(s)\beta_j\psi_0\right)ds\no\\
  & & -i\ve e^{i\la_0t}\ \left(\psi_0,\  W(t)\phi_0 (t)\right) \nonumber \\
  & & - i\ve e^{i\la_0 t}\ \left(\psi_0,\  W(t)\phi_2 (t)\right). 
\label{eq:Eexpand}
\ea
 Here, 
 $\phi_0$ and $\phi_2$ are given in (\ref{eq:intphideqn}).      
 \end{prop}
Although the proposition is stated for $t>0$, an analogous proposition with $-\ve^2\Gamma$ replaced by $\ve^2\Gamma$ holds for $t<0$. The modification required to treat $t<0$ is indicated in the proof.
\medskip
\begin{rmk}
\nit (1) The point of (\ref{eq:Aeqn3}) is that the source 
of damping,   
$\Gamma>0$, which arises due to the coupling of the discrete bound
state to the continuum modes by the almost periodic perturbation, is made
explicit. Note that $\Re\rho(t)$ is of order $\ve^2|||W|||^2$ as the first two terms of $\rho(t)$ are pure imaginary 
inducing only a phase shift in the solution, $A(t)$, while the last one is
of the same order as the damping, and may compete with it. A key 
point of our analysis is to assess the contribution of this last term in 
(\ref{eq:rhoexpansion}).

\nit (2) The leading order part
 of equation (\ref{eq:Aeqn3}) is the analogue of the dispersive
normal form derived in \cite{kn:rdamping} for a class of nonlinear
dispersive wave
equations.
\end{rmk}

\noindent
\underline{Proof of Proposition \ref{pr:Aequation}} 

Using the expression for $W(t)$ in (\ref{eq:generalW1}), which is a uniform convergent 
series with respect to $t\in\R$, and the definition 
$A(t)=e^{i\lambda_0t}a(t)$, we get from (\ref{eq:intphideqn})
\ba
\phi_1(t) &=&\  -{i\ve\over 2}\int^t_0 e^{-iH_0(t-s)} 
 e^{-i\la_0s }A(s) \Pc \sum_{j\in\Z}e^{-i\mu_js}\ \beta_j\psi_0\ ds \no\\ 
          &=&\  -{i\ve\over 2}\sum_{j\in\Z} \int^t_0 e^{-iH_0(t-s)} 
 e^{-i(\la_0 + \mu_j)s }A(s) \Pc  \beta_j\psi_0\ ds \label{eq:phidexpand}
 \ea
We would like to integrate by parts each of the integrals in the above sum. We
cannot proceed directly since the resolvents of $H_0$ in $\la_0+\mu_j,\ 
j\in\Z,$ would
appear and hypothesis {\bf (H5)} implies that some of the $\la_0+\mu_j,\ 
j\in\Z,$ are in the spectrum of $H_0$. Instead we regularize $\phi_1$ by
defining:
\be
\phi_1^\eta (t)=\ -{i\over 2}\ve\sum_{j\in\Z} \int^t_0 e^{-iH_0(t-s)} 
 e^{-i(\la_0 + \mu_j+i\eta )s }A(s) \Pc  \beta_j\psi_0\ ds \label{eq:Kepsilon}
 \ee
for $\eta$ positive and arbitrary and $t>0$. Note that $\phi_1(t)=
\lim_{\eta\searrow 0}\phi_1^\eta (t)$ uniformly with respect to $t$ on compact intervals.

Now, integration by parts for each integral in expression (\ref{eq:Kepsilon})
and letting $\eta$ tend to zero from above gives the following expansion of
$\left(\psi_0,W(t)\phi_1(t)\right)$ :
\ba
\lefteqn{\left(\psi_0,W(t)\phi_1(t)\right) = \left(W(t)\psi_0,\ - {\ve\over 2} \ e^{-i\lambda_0 t}\sum_{j\in\Z}e^{-i\mu_jt} A(t) 
  (H_0 - \la_0 - \mu_j- i0)^{-1} \Pc  \beta_j\psi_0\right)} \no \\
&+&\left(W(t)\psi_0,\  {\ve\over 2} A(0)\sum_{j\in\Z}e^{-iH_0t} (H_0 - \la_0 - \mu_j -i0)^{-1} \Pc  
\beta_j\psi_0\right)
\label{eq:Kresexpansion}\\
&+&\left(W(t)\psi_0,\  {\ve\over 2}\sum_{j\in\Z}\int_0^t e^{-iH_0(t-s)} 
 (H_0 - \la_0 - \mu_j- i0)^{-1}\Pc e^{-i(\la_0+\mu_j)s} \D_sA(s) \beta_j\psi_0
 ds\right). \no 
\ea
The definition of the singular operators in the above computation is given in section \ref{se:sop}. The choice of regularization, $+i\eta$, in (\ref{eq:Kepsilon}) ensures
that the latter two terms in the expansion of $\phi_1$,
(\ref{eq:Kresexpansion}), decay dispersively as $t \to +\infty$; see
hypothesis {\bf (H3)} and section \ref{se:localdecay}. For $t<0$, we replace
$+i\eta$ with $-i\eta$ in  (\ref{eq:Kepsilon}).

To further expand the first series in (\ref{eq:Kresexpansion}) we use the 
identities (\ref{eq:distrid}). The proof of Proposition \ref{pr:Aequation} is now completed by substitution 
 of (\ref{eq:distrid}) in the expansion (\ref{eq:Kresexpansion}) for 
 $\phi_1$ and of the result into the second term of the sum in (\ref{eq:Aeqn2}).
\ \ \ []
\bigskip

In the next sections we  estimate the remainder terms in (\ref{eq:intphideqn})
and 
(\ref{eq:Aeqn3}). %
\section{Estimates on the bound state amplitude}\label{se:ode}

Our strategy is as follows.
Equations (\ref{eq:intphideqn}) and (\ref{eq:Aeqn3}) comprise a dynamical
 system governing  
  $\phi_d(t)$ and $a(t)=A(t)e^{-i\la_0t}$, the solution of which is equivalent to the 
 original equation (1.1).  
  In this and in the following section we derive a coupled system of
  estimates for $A(t)$ and $\phi_d(t)$. 
This section is focused on obtaining estimates for the bound state amplitude
$A(t)$ in terms of $\phi_d(t)$, while the following section is focused on 
obtaining dispersive estimates for $\phi_d(t)$ in terms of $A(t)$. We treat only the case $t>0$ since the modifications for the case $t<0$ are obvious.
The coupled system of estimates shows that 
  $A(t)$ decays in time, provided $\phi_d(t)$ is dispersively decaying
  and vice-versa.  We exploit the assumed smallness of the perturbation 
  $\ve W$ to "close" the resulting inequalities, and prove the decay
  of both $A(t)$ and $\phi_d(t)$. 

The main difference from the strategy employed in \cite{kn:nh} for the
  estimation  of the  bound state amplitude is related to the presence of
{\it infinitely many} frequencies in the perturbation $W(t)$. In 
  particular one can have an {\it accumulation of resonances}
 in the continuous spectrum of $H_0$.
 We have two strategies for obtaining  estimates for $A(t)$ which 
 correspond to the use of hypotheses {\bf (H1)-(H5)} 
(Theorems \ref{th:twofive} and \ref{th:twosix}) or hypotheses {\bf (H1)-(H6)}
 (Theorem \ref{th:twotwo}). These strategies revolve around estimation of 
$\Re \int_0^t\ \rho(s)\ ds$, where $\rho$ is given by (\ref{eq:rhoexpansion}). {\bf (H6)}, which controls certain 
"small divisors'' which arise from the clustering of frequencies, ensures that 
\be \Re \int_0^t\ \rho(s)\ ds\ \le\ C\ \ve^2|||W|||^2.\no\ee This, in turn,
implies that the contribution of $\rho(t)$ in the size of $A(t)$ is of order $\ve^2|||W|||^2$.
Without hypothesis {\bf (H6)} we carefully decompose
 $\rho(t)$ as
$$ \rho(t)\ =\ \ve^2\sigma(t)\ +\ \eta(t),$$
where $\sigma(t)$ is a real almost periodic function with mean, $M(\sigma)$, zero
and $\Re\ \int_0^t\ \eta(s)\ ds\ \le\  C \ve^2 ||W|||^2$. As in the previous case,
 the contribution of the $\eta(t)$ in the size of $A(t)$ is of order $\ve^2|||W|||^2$. On the other hand, 
$\sigma(t)$ competes with the damping term $\ve^2\Gamma$ in equation 
(\ref{eq:Aeqn3}), but being oscillatory (i.e of mean zero) and of the 
same size as the damping it allows the latter to eventually dominate. 

As the above discussion suggests it is simplest to start by assuming {\bf (H6)} 
to get sharper estimates on $A(t)$ (Theorem \ref{th:twotwo}) and then to relax this assumption (Theorem \ref{th:twosix}).
We begin with a simple Lemma which we shall use in a number of
  places in this and in the next section.

  \begin{lem}  Let $\alpha>1$.
  \be
  \int_0^t\ \lan t-s\ran^{-\alpha}\ \lan s\ran^{-\beta}\ ds\ \le
  C_{\alpha,\beta}\ \lan t\ran^{-\min{(\alpha,\beta)}}
  \nn\ee
  \end{lem}

  \nit\underline{Proof:} The bound is obtained by viewing the integral
  as decomposed into 
   a part over $[0,t/2]$ and the part over $[t/2,t]$.  We estimate
  the integral over $[0,t/2]$  by 
    bounding $\lan t-s\ran^{-\alpha}$ by its value at $t/2$ and explicitly
	computing the remaining integral. The integral over $[t/2,t]$ is
	computed by bounding $\lan s\ran^{-\beta}$ by its value at $t/2$ and
	again computing explicitly the remaining integral. Putting the two
	estimates together yields the lemma.

We now turn to the estimate for $A(t)$ in terms of the dispersive norm of 
 $\phi_d(t)$ and local decay estimates for $e^{-iH_0 t} \Pc  (H_0)$. 

\bigskip

 \subsection{Estimates for  $A(t)$
  under the hypotheses of Theorem \ref{th:twotwo}}\label{sse:A}
\medskip

\begin{prop}\label{pr:odeest} Suppose {\bf (H1)-(H6)} hold. Then
$A(t)$, the solution of (\ref{eq:Aeqn3}), can be expanded as:
\ba
  A(t)\ &=&\ e^{\int^t_0\rho(s)ds} \left( e^{-\ve^2\Gamma t} A(0) + R_A(t) \right)
  \label{eq:prop4a}\\
  R_A(t)\ &=&\ \int_0^t\ e^{-\ve^2\Gamma(t-\tau)}\ \tilde E(\tau)\ d\tau,
  \label{eq:prop4aa}
\ea
where $\tilde E(t)$ is given in (\ref{eq:Eexpand}) and  (\ref{eq:tE}).
For any $\alpha>1$, there exists a $\delta >0$ such that $R_A(t)$ satisfies the estimates for 
 $T>2(\ve^2\Gamma)^{-\alpha}$, 
\ba
\sup_{2(\ve^2\Gamma)^{-\alpha}\le t\le T} \lan t\ran^{r_1} \ |R_A(t)| 
 &\le& C_1e^{-(\ve^2\Gamma)^{-\delta}} \sup_{0\le\tau\le \M}|E(\tau)|\no\\
&+&\ 
  C\ \Gamma^{-1}\ \sup_{\M\le \tau\le T}
  \left(\lan\tau\ran^{r_1}|E(\tau)|\right),
  \label{eq:prop4b}\\
  \sup_{0\le t\le 2(\ve^2\Gamma)^{-\alpha}} \lan t\ran^{r_1} \ |R_A(t)|
   &\le&\ D\ (\ve^2\Gamma)^{-\alpha (r_1+1)}\ \sup_{0\le \tau\le 2(\ve^2\Gamma)^{-\alpha} }\ 
   |E(\tau)| \label{eq:prop4c}
\ea
\end{prop}
\noindent
\underline{Proof.}
To prove (\ref{eq:prop4b}) we begin with (\ref{eq:Aeqn3}). Let
\be
\tilde{A}(t) \equiv e^{- \int^t_0 \rho(s)ds} A(t).\label{eq:tAdef}
\ee
Then, $\tilde{A}$ satisfies the equation
\begin{eqnarray}
\D_t\tilde{A} &=& -\ve^2\Gamma \tilde{A} + \tilde{E}(t) \label{eq:tAeqn1}\\
\tilde{E}(t) &\equiv& e^{-\int^t_0 \rho(s)ds} E(t). \label{eq:tE}
\end{eqnarray}
Solving (\ref{eq:tAeqn1}) we get
\ba
\tilde{A}(t)\ &=&\ e^{-\ve^2\Gamma t} \tilde{A}(0)\ +\
 \int^t_0 e^{-\ve^2\Gamma(t-s)} \tilde{E}(s)\ ds
\label{eq:tAinteqn}\\
&\equiv&\ e^{-\ve^2\Gamma t} \tilde{A}(0)\ +\ R_A(t).
\ea

Below, in Proposition \ref{le:rhobounds} we show 
  that the real part of the integral of $\rho(t)$ is uniformly bounded and
of order ${\cal O}(\ve^2|||W|||^2)$, for $t\ge 0$.  Therefore, for some $C>0$, we have 
by (\ref{eq:tAdef}) and
(\ref{eq:tE})
\ba
 C^{-1} |\tilde{A}(t)| &\le& |A(t)| \le C |\tilde{A}(t)| \label{Abounds}\\
 C^{-1} |\tilde{E}(t)| &\le& |E(t)| \le C |\tilde{E}(t)| \label{Ebounds}
\ea  
 Consequently, it is sufficient to estimate 
  $\tilde{A}(t)$, in terms of $\tilde{E}(t).$  

\begin{rmk} 
 Estimates of 
   $R_a(t)$, which appears in the statement of Theorem \ref{th:twotwo},
   are  related to those for  $R_A(t)$ via :
   \be\label{eq:rararel1}
   R_a(t)\ =\ e^{-i\la_0t+\int_0^t\rho(s)\ ds} \ R_A(t)-
\left(1-e^{\Re\int_0^t\rho(s)ds}\right)e^{-\ve^2\Gamma t}a(0).\ee
Hence, by Proposition \ref{le:rhobounds},
\be \label{eq:rararel12} 
\left|R_a(t)\right|\ \le\ C\ \left|R_A(t)\right|+{\cal O}(\ve^2|||W|||^2)\ee
\end{rmk}

 From (\ref{eq:tAinteqn}) we have for any $M>0$: 
\ba
|\tilde{A}(t)|\ &\le& |A(0)|e^{-\ve^2\Gamma t} + 
 \int^M_0 e^{-\ve^2\Gamma(t-s)}  |\tilde{E}(s)| ds +
\int^t_M e^{-\ve^2\Gamma(t-s)} |\tilde{E}(s)|\ ds\no\\
&=&\ |A(0)|e^{-\ve^2\Gamma t}\ +\ I_1(t)\ +\ I_2(t).
 \label{eq:tAM}
\ea
Set 
 $$M=(\ve^2\Gamma)^{-\alpha},\ \ \alpha>1.$$
We now estimate the terms $I_1(t)$ and $I_2(t)$
in (\ref{eq:tAM}) for $2\M\le t\le T$.
\ba
\lan t\ran^{r_1}\ I_1(t)\ &=&\ \lan t\ran^{r_1}\ \int^M_0 e^{-\ve^2\Gamma(t-s)}
|\tilde{E}(s)| ds\no\\
&\le&\ \lan t\ran^{r_1}e^{-{1\over2}\ve^2\Gamma t }
 \cdot \int^M_0 e^{-\ve^2\Gamma ({1\over2}t-s)}\ ds\
\cdot \sup_{0\le\tau\le\M} |\tilde E(\tau)|\no\\
&\le&\ 
\sup_{2\M\le t\le T}\left(\lan t\ran^{r_1}e^{-{1\over2}\ve^2\Gamma t}\right) \cdot
 C(\ve^2\Gamma)^{-1} \cdot 
  \sup_{0\le\tau\le\M} |\tilde E(\tau)|\no\\
 &\le&\ Ce^{-(\ve^2\Gamma)^{-\delta}}\ \sup_{0\le\tau\le\M} |\tilde E(\tau)|,
\label{eq:I1est1}
\ea 
for some $\delta>0$. Therefore, 
\be
\sup_{2\M\le t\le T}\left(\lan t\ran^{r_1}\ I_1(t)\right)
\ \le \  Ce^{-(\ve^2\Gamma)^{-\delta}}\ \sup_{0\le\tau\le\M} |\tilde E(\tau)|
\label{eq:I1est2}
\ee
We estimate $I_2(t)$ on the interval $2\M\le t\le T$ as follows:
\be
\lan t\ran^{r_1}\ I_2(t)\ \le\  
 \lan t\ran^{r_1}\  \int_{\M}^t e^{-\ve^2\Gamma(t-s)} 
  \lan s\ran^{-r_1} \ ds\   
   \sup_{\M\le\tau\le T}\left(\lan\tau\ran^{r_1}\tilde{E}(\tau)\right)
   \label{eq:I2est1}
\ee
The integral is now bounded above using the estimate
\be
\lan t\ran^{r_1}\ \int_{\M}^t\ e^{-\ve^2\Gamma(t-s)}\ \lan s\ran^{-r_1}\ ds\ 
    \le\ C(\ve^2\Gamma)^{-1},\ t\ge2\M.
\label{eq:integralestimate}
\ee

This gives
\be
\sup_{2(\ve^2\Gamma)^{-\alpha}\le t\le T}\lan t\ran^{r_1}\ I_2(t)\ 
 \le\  C(\ve^2\Gamma)^{-1}\ 
\sup_{\M\le\tau\le T}\left(\lan\tau\ran^{r_1}\tilde{E}(\tau)\right)
\label{eq:I2est2}
\ee

Assembling the estimates (\ref{eq:I1est2}) and (\ref{eq:I2est2})
yields estimate (\ref{eq:prop4b}) of Proposition \ref{pr:odeest} provided that 
(\ref{Abounds}) and (\ref{Ebounds}) hold. 
 Estimate (\ref{eq:prop4c}) is a simple consequence of 
  the definition of $R_A(t)$.
  
Thus it remains to prove (\ref{Abounds}) and (\ref{Ebounds}).
By (\ref{eq:tAdef}) and 
 (\ref{eq:tE}) it is necessary
and sufficient to verify the following proposition:

\begin{prop}\label{le:rhobounds} 
Assume hypotheses {\bf (H1)-(H6)}. 
If $\rho$ is given by (\ref{eq:rhoexpansion}) 
then 
\be
\Re\int_0^t\ \rho(s)\ ds \le C\ve^2|||W|||^2 \label{eq:exrhobounds},\ t\ge 0,
\ee
for some constant $C$ depending on ${\cal C},\ r_1$ and $\xi$; see {\bf (H6)}.
\end{prop}

\nit\underline{Proof of Proposition \ref{le:rhobounds}:} Using the estimates
(\ref{eq:rezbound}) and (\ref{eq:PVbound}) we can infer that, $\rho(t)$, given by (\ref{eq:rhoexpansion}) 
 is a series which converges uniformly on any compact subset of $\R$. For each
fixed $t$, it can therefore be integrated term by term to give:
\ba
\Re\int_0^t\rho(s)ds &=& {\ve^2\over4}\Re\ i\ \sum_{j,k\in\Z,j\neq k}\int_0^t
e^{i(\mu_k-\mu_j)s}\left(\beta_k\psi_0, 
  (H_0-\lambda_0-\mu_j-i0)^{-1}\Pc\beta_j\psi_0\right) ds \no\\
 &=& {\ve^2\over4}\sum_{j,k\in\Z,j\neq k}\Re
 \frac{e^{i(\mu_k-\mu_j)t}-1}{\mu_k-\mu_j}
\left(\beta_k\psi_0,(H_0-\lambda_0-\mu_j-i0)^{-1}\Pc\beta_j\psi_0\right)
\label{eq:rhobounds1}
\ea
Define 
$$
\tilde{\rho}_{j,k}\equiv 
 \frac{e^{i(\mu_k-\mu_j)t}-1}{\mu_k-\mu_j}
\left(\beta_k\psi_0,(H_0-\lambda_0-\mu_j-i0)^{-1}\Pc\beta_j\psi_0\right).
$$
Then (\ref{eq:rhobounds1}) can be expressed as: 
\be
\Re\int_0^t\rho(s)ds={\ve^2\over4}\sum_{j,k\in\Z,j\neq k}\Re\tilde{\rho}_{j,k}=
{\ve^2\over8}\sum_{j,k\in\Z,j\neq k}\Re (\tilde{\rho}_{j,k}+\tilde{\rho}_{k,j}).
\label{eq:sharp1}\ee
Now, since
\ba
\tilde\rho_{k,j}&=& -\frac{e^{-i(\mu_k-\mu_j)t}-1}{\mu_k-\mu_j}
\left(\beta_j\psi_0,(H_0-\lambda_0-\mu_k-i0)^{-1}\Pc\beta_k\psi_0\right)\no\\
&=&-\overline{\frac{e^{i(\mu_k-\mu_j)t}-1}{\mu_k-\mu_j}
\left(\beta_k\psi_0,(H_0-\lambda_0-\mu_k+i0)^{-1}\Pc\beta_j\psi_0,\right)}\no
\ea
we have
\be
\Re(\tilde\rho_{j,k}+\tilde\rho_{k,j})=
  \Re\frac{e^{i(\mu_k-\mu_j)t}-1}{\mu_k-\mu_j}
\left(\beta_k\psi_0,(H_0-\lambda_0-\mu_j-i0)^{-1}-(H_0-\lambda_0-\mu_k+i0)^{-1}\Pc\beta_j\psi_0\right)
\label{eq:sharp2}\ee
Moreover, by (\ref{eq:distrid}) we can infer 
$$
\Re(\tilde\rho_{j,k}+\tilde\rho_{k,j})=\Re\left(e^{i(\mu_k-\mu_j)t}-1\right)\rho_{j,k}
+2\Im\left(e^{-i(\mu_k-\mu_j)t}-1\right)\delta_{j,k},\no
$$
where, for $j\neq k\in\Z$,
\be
\rho_{j,k}\equiv\frac{1}{\mu_k-\mu_j}\left(\beta_k\psi_0,
(H_0-\lambda_0-\mu_j-i0)^{-1}-(H_0-\lambda_0-\mu_k-i0)^{-1}\Pc\beta_j\psi_0
\right),\label{eq:defrho}\ee
and for $j\neq k,\ j\in\Z,\ k\in\res$,
\be
\delta_{j,k}\equiv\frac{\pi}{\mu_k-\mu_j}\left(\beta_j\psi_0,
\delta(H_0-\la_0-\mu_k)\beta_k\psi_0\right).\label{eq:defdelta}
\ee

Thus, by (\ref{eq:sharp1}) and (\ref{eq:sharp2})
\be
\Re\int_0^t\rho(s)ds = {\ve^2\over8}\sum_{j,k\in\Z,j\neq k}\Re 
\left({e^{i(\mu_k-\mu_j)t}-1}\right)\rho_{j,k}
+{\ve^2\over4}\sum_{k\in\res ,k\neq j\in\Z}\Im 
\left({e^{-i(\mu_k-\mu_j)t}-1}\right)\delta_{j,k}
.\label{eq:rhobounds}
\ee
We now derive a uniform bound for $\Re\ \int_0^t\rho(s)\ ds$.

Estimating the modulus of the above sum, we have for any $t$:
\be
\left|\ \Re\ \int_0^t\rho(s)\ ds\ \right|\ \le\
  {\ve^2\over4}\sum_{j,k\in\Z,j\neq k}
\left|\rho_{j,k}\right|+{\ve^2\over2}\sum_{k\in\res ,k\neq j\in\Z}
\left|\delta_{j,k}\right| .\label{eq:rhofinal}
\ee
By {\bf (H6)}, 
\be
\sum_{k\in\res ,k\neq j\in\Z}\ |\delta_{j,k}|\ \le\ \pi \xi\ |||W|||^2.
\label{eq:bounda}\ee

We now bound the first term in (\ref{eq:rhofinal}). This requires an estimate
of:
$$\left|\rho_{j,k}\right|=\left|\frac{1}{\mu_k-\mu_j}\left(\beta_k\psi_0,
(H_0-\lambda_0-\mu_j-i0)^{-1}-(H_0-\lambda_0-\mu_k-i0)^{-1}\Pc\beta_j\psi_0
\right)\right|,$$
for $j\neq k\in\Z$. We rely on the hypothesis {\bf (H3b)} (singular local decay estimate (\ref{eq:singularldest}) ), which imply smoothness of the resolvent of $H_0$ near 
 accumulation points in $\spectrum$ of the set
$\{\la_0+\mu_j\}_{j\in\Z}$. 

In order to treat both $\la_0+\mu_j\in\spectrum$ and 
$\la_0+\mu_j\notin\spectrum$ case simultaneously we regularize $\rho_{j,k}$:
\ba
\lefteqn{\rho_{j,k}^{\eta}\equiv \frac{1}{\mu_k-\mu_j}}\no\\
& & \left(\beta_k\psi_0,
(H_0-\lambda_0-\mu_j-i\eta)^{-1}-(H_0-\lambda_0-\mu_k-i\eta)^{-1}\Pc\beta_j\psi_0
\right).\label{eq:rhoreg}
\ea
Clearly $\rho_{j,k}=\lim_{\eta\searrow 0}\rho_{j,k}^{\eta}$

Now by the standard resolvent formula we have:
$$
\rho_{j,k}^{\eta}=\left(\beta_k\psi_0,
(H_0-\lambda_0-\mu_k-i\eta)^{-1}(H_0-\lambda_0-\mu_j-i\eta)^{-1}\Pc\beta_j\psi_0
\right).
$$
Thus, using the singular local decay estimate {\bf (H3b)}, we get: 
\ba
|\rho_{j,k}| &=& \left|\lim_{\eta\searrow 0}\int_0^\infty 
\left(\beta_k\psi_0,e^{-i
(H_0-\lambda_0-\mu_k-i\eta)s}(H_0-\lambda_0-\mu_j-i\eta)^{-1}\Pc\beta_j\psi_0
\right)ds\right|\no\\
&\leq & \lim_{\eta\searrow 0}\int_0^\infty e^{-\eta s}\left|\left(\wpl\beta_k\psi_0,\wmi e^{-i
H_0s}(H_0-\lambda_0-\mu_j-i\eta)^{-1}\Pc\wmi\wpl\beta_j\psi_0
\right)\right|ds\no\\
&\leq & \|\wpl\beta_k\|\|\wpl\beta_j\|\int_0^\infty\|\wmi e^{-i
H_0s}(H_0-\lambda_0-\mu_j-i0)^{-1}\Pc\wmi\|ds\no\\
&\leq & {\cal C}\|\wpl\beta_k\|\|\wpl\beta_j\|\int_0^\infty \lan s\ran^{-r_1}ds\no\\
&\leq & C \|\wpl\beta_k\|\|\wpl\beta_j\|,\label{eq:rhojkbounds}
\ea
for some constant $C$ depending on ${\cal C}$ and $r_1$.
 Summing on $j,k\in\Z,\ j\ne k$ yields:
\be
\sum_{j,k\in\Z,\ j\ne k}\ |\rho_{j,k}|\ \le\ C\ |||W|||^2, 
\label{eq:boundb}
\ee
for some $C>0$; see (\ref{eq:betanorm}).
Use of the bounds (\ref{eq:bounda}) and (\ref{eq:boundb}) in 
 (\ref{eq:rhofinal}) gives
$$
\left|\ \Re\ \int_0^t\ \rho(s)\ ds\ \right|\ \le\ C\ \ve^2|||W|||^2 \no
$$
for some constant $C$ depending on ${\cal C},\ r_1$ and $\xi$.

This completes the proof of 
  Proposition \ref{le:rhobounds} and therewith Proposition
\ref{pr:odeest}. \ \ \ \ []
\bigskip


\subsection{Estimates for $A(t)$ under the hypotheses of Theorem \ref{th:twofive}}\label{sse:B}

In this subsection  we work under the hypotheses of Theorem \ref{th:twofive}. In particular,
we drop hypothesis {\bf (H6)}.
 {\it We shall reuse the notation $\tilde A$ and $\tilde E$ for functions 
 which are different from but related to  those defined in section 5.1.}
\medskip

\begin{prop}\label{pr:odeest1} Suppose {\bf (H1)-(H5)} hold. Then
$A(t)$, the solution of (\ref{eq:Aeqn3}), can be expanded as:
\ba
  A(t)\ &=&\ e^{\int^t_0\eta(s)ds} \left( e^{-\ve^2(\Gamma t-\int_0^t\sigma
  (s)ds)} A(0) + R_A(t) \right)
  \label{eq:prop4a1}\\
  R_A(t)\ &=&\ \int_0^t\ e^{-\ve^2\Gamma(t-\tau)+
 \ve^2\int_\tau^t\sigma (s)ds}\ \tilde E(\tau)\ d\tau,
  \label{eq:prop4aa1}
\ea
where 
\be
\sigma(t)\equiv -\frac{\pi}{4}\Re \sum_{j\in\res ,j\neq k\in\Z}e^{i(\mu_k-\mu_j)t}
\left(\ \beta_k\psi_0,\delta(H_0-\la_0-\mu_j)\beta_j\psi_0\right).
 \label{eq:sigma}
\ee
 is a real almost periodic function with
 mean $M(\sigma)=0$, $\eta$ in (\ref{eq:varrho}) is a function whose real part
has a bounded time-integral of order ${\cal O}(\ve^2|||W|||^2)$ and 
 $\tilde E(t)$ is given in (\ref{eq:tE1}), see also (\ref{eq:Eexpand}). 
For any $\alpha>1$, there exists $\delta >0$ such that $R_A(t)$ satisfies the estimates:
\ba
\sup_{2(\ve^2\Gamma/2)^{-\alpha}\le t\le T} \lan t\ran^{r_1}|R_A(t)| 
 &\le & C_1e^{-(\ve^2\Gamma/2)^{-\delta}} \sup_{0\le\tau\le \MM}|E(\tau)|\no\\
&+& 
  C(\ve^2\Gamma)^{-1}\sup_{\MM\le \tau\le T}
  \left(\lan\tau\ran^{r_1}|E(\tau)|\right) ,
  \label{eq:prop4b1}\\
  \sup_{0\le t\le 2(\ve^2\Gamma/2)^{-\alpha}} \lan t\ran^{r_1} \ |R_A(t)|
   &\le &\ D\ (\ve^2\Gamma/2)^{-\alpha (r_1+1)}\ \sup_{0\le \tau\le 2(\ve^2\Gamma/2)^{-\alpha}}\ 
   |E(\tau)|. \label{eq:prop4c1}
\ea
\end{prop}
\noindent 
\underline{Proof.}
 As in the previous subsection we begin with the equation for $A(t)$:
\be
  \D_tA(t)\ =\ \left( \rho(t)\ -\ \ve^2\Gamma\ \right) A(t) + E(t), 
\label{eq:Aeqn3a}\ee
where $\rho(t)$ and $E(t)$ are given by (\ref{eq:rhoexpansion}-\ref{eq:Eexpand}).
In the previous section we transformed away the term 
  $\rho(t)A(t)$ using the "integration factor": $\exp(\int_0^t\rho(s)\ ds)$.
 Under the current hypotheses, this can't be done because without {\bf (H6)}
$\Re \int_0^t\rho(s)\ ds$
 may be unbounded as $t\rightarrow\infty$, which could cause the estimates
 (\ref{Abounds}-\ref{Ebounds}) to break down.  Instead, we proceed by a more refined analysis of $\rho(t)$, which we 
 now outline. 
 
We express $\rho(t)$ as $\rho(t) = \ve^2\sigma(t) + \eta(t)$, where $\eta(t)$ has a
 time integral whose  
 real part can be bounded by the estimates of section 5.1 and a part, $\ve^2\sigma(t)$
 which is almost periodic and of mean zero.  Using this decomposition of $\rho(t)$ we write 
  (\ref{eq:Aeqn3a}) as
$$
\D_t A(t)\ =\ \left[ -\ve^2\Gamma\ +\ \ve^2\sigma(t)\ + \eta(t) \right]A(t)
 \ +\ E(t).
$$
 Next  introduce the change of variables
\be
\tilde A(t)\ \equiv\ e^{-\int_0^t\eta(s)\ ds}\ A(t)
\label{eq:tAdef1}\ee
and obtain a reduction to 
\ba
\D_t\tilde{A} &=& \left[\ -\ve^2\Gamma\ +\ \ve^2\sigma(t)\ \right] \tilde{A} 
\ +\  \tilde{E}(t)
\label{eq:tAeqn11}\\
\tilde{E}(t) &\equiv& e^{- \int^t_0 \eta(s)ds} E(t). 
 \label{eq:tE1}
\ea

With this strategy in mind we now 
 proceed to derive the decomposition of $\rho(t)$. We are mostly interested in
 its real part, so we start with it.

\ba
\Re\rho(t)&=&\Re{i\ve^2\over4}\sum_{j,k\in\Z,j\neq k}e^{i(\mu_k-\mu_j)t}\left(\beta_k\psi_0,
  (H_0-\lambda_0-\mu_j-i0)^{-1}\Pc\beta_j\psi_0\right)\no\\
  &=&-{\ve^2\over4}\Im\sum_{j,k\in\Z,j\neq k}e^{i(\mu_k-\mu_j)t}\left(\beta_k\psi_0,
  (H_0-\lambda_0-\mu_j-i0)^{-1}\Pc\beta_j\psi_0\right)\no\\
  &\equiv&-{\ve^2\over4}\sum_{j,k\in\Z,j\neq k}\Im\eta_{j,k}\no\\
  &=&{\ve^2\over8}\sum_{j,k\in\Z,j\neq
  k}\Im\left(\eta_{j,k}+\eta_{k,j}\right).\label{eq:exprrho}
\ea
In a manner similar to the derivation of (\ref{eq:sharp2}) from (\ref{eq:sharp1}) we find 
\be
\Im \eta_{k,j}=\Im e^{i(\mu_k-\mu_j)t}\left(\beta_k\psi_0, (H_0-\lambda_0-\mu_k+i0)^{-1}\Pc
  \beta_j\psi_0\right).\label{eq:expretakj}
\ee
Using (\ref{eq:distrid}) in (\ref{eq:expretakj}) and then replacing it in (\ref{eq:exprrho}) we get 
\ba
\lefteqn{\Re \rho(t)\ =\ {\pi\over4}\ve^2\Re \sum_{k\in\res ,k\neq j\in\Z} e^{i(\mu_j-\mu_k)t}
 \left(\beta_j\psi_0,
  \delta(H_0-\lambda_0-\mu_k)\beta_k\psi_0\right)}\no\\
  &-&{1\over8}\ve^2 \Im\ \sum_{j,k\in\Z,j\neq k} e^{i(\mu_k-\mu_j)t}
 \left(\beta_k\psi_0,
  \left[ (H_0-\lambda_0-\mu_j-i0)^{-1}-(H_0-\lambda_0-\mu_k-i0)^{-1}
 \right]\Pc\beta_j\psi_0\right)\no\\
 &=&\ \Re\ \eta(t)+\ve^2\sigma(t).\no
\ea
Therefore,
\ba \rho(t)\ &=&\ \Re\rho(t)\ +\ i\Im\rho(t)\no\\
             &=&\ \eta(t)\ +\ \ve^2\sigma(t)\label{eq:exprrho1}\ea
where
\ba
\eta(t)&=&i\ \Im\ \rho(t)-{1\over8}\ve^2 \Im\ 
 \sum_{j,k\in\Z,j\neq k} e^{i(\mu_k-\mu_j)t}\no\\  
 & &\left(\beta_k\psi_0,
  \left[ (H_0-\lambda_0-\mu_j-i0)^{-1}-(H_0-\lambda_0-\mu_k-i0)^{-1}\Pc\right]
  \beta_j\psi_0\right) \label{eq:varrho}\\ 
\sigma(t)&=& 
 -\frac{\pi}{4}\Re \sum_{j\in\res ,j\neq k\in\Z}e^{i(\mu_k-\mu_j)t}
\left(\ \beta_k\psi_0,\delta(H_0-\la_0-\mu_j)\beta_j\psi_0\right),
 \no
\ea
see also (\ref{eq:sigma}).

Note that 
$\Re\ \int_0^t\eta(s)ds$
is uniformly bounded in $t$. To see this, recall the definition of
$\rho_{j,k}$ in Lemma \ref{le:rhobounds} (see (\ref{eq:defrho})):
$$\rho_{j,k}\equiv \frac{1}{\mu_k-\mu_j}\left(\beta_k\psi_0,
\left[ (H_0-\lambda_0-\mu_j-i0)^{-1}-(H_0-\lambda_0-\mu_k-i0)^{-1}\right]
  \Pc\beta_j\psi_0
\right).$$
By (\ref{eq:rezbound}), $\Re\ \eta(t)$ given by (\ref{eq:varrho}), converges uniformly on $t\in\R$. Therefore, 
 for each $t\in\R$ we may 
integrate the series term by term to obtain
\be
\Re\ \int_0^t\eta(s)ds={1\over8}\ve^2\sum_{j,k\ j\neq k}\Re
\left(e^{i(\mu_k-\mu_j)t}-1\right)\rho_{j,k}.\label{eq:huf}
\ee
Moreover the modulus of the right hand side in (\ref{eq:huf}) is less or equal
than ${1\over4}\ve^2\sum_{j,k\ j\neq k}|\rho_{j,k}|$ which by
(\ref{eq:rhojkbounds}) is bounded by $C\ve^2 |||W|||^2$ for some constant $C$
depending only on ${\cal C}$ and $r_1$. Note that we derived (\ref{eq:rhojkbounds})
by using only hypothesis {\bf (H3b)} and not relying on {\bf (H6)}.

Thus we have
\be\label{eq:varrhobounds1}
\Re\ \int_0^t\eta(s)ds\le C\ve^2 |||W|||^2.
\ee

To summarize, we have split $\rho(t)$ into
$$
\rho(t)=\eta(t)\ +\ \ve^2\sigma(t),
$$
such that (\ref{eq:varrhobounds1}) is valid. If we now define $\tilde{A}$
as in (\ref{eq:tAdef1}) 
then, by (\ref{eq:Aeqn3}) $\tilde{A}$ satisfies the equation 
 (\ref{eq:tAeqn11}).
Solving (\ref{eq:tAeqn11}) we get
\ba
\tilde{A}(t)\ &=&\ e^{-\ve^2\Gamma t+\ve^2\int_0^t\sigma(s)ds} \tilde{A}(0)\ +\
 \int^t_0 e^{-\ve^2\Gamma(t-\tau)+\ve^2\int_\tau^t\sigma(s)ds} \tilde{E}(s)\ d\tau
\no\\
&\equiv&\ e^{-\ve^2\Gamma t+\ve^2\int_0^t\sigma(s)ds} \tilde{A}(0)\ +\ R_A(t).
\label{eq:tAinteqn1}
\ea

From (\ref{eq:tE1}) and (\ref{eq:varrhobounds1}) it is sufficient to estimate 
$R_A(t)$, in terms of $\tilde{E}(t)$.

\begin{rmk} The estimates of $R_a(t)$
which appears in the statement of Theorem \ref{th:twosix}, are related to
those for $R_A(t)$ via:
\be 
R_a(t)\ =\ e^{-i\la_0t+\int_0^t\eta(s)ds}R_A(t)+
\left(1-e^{\ve^2(\int_0^t\sigma(s)ds-\gamma t)+\Re\int_0^t\eta(s)ds}\right)e^{-\ve^2(\Gamma-\gamma )t}a(0).
\label{eq:rararelation}
\ee
\end{rmk}

Before we estimate $R_A(t)$, we  review
some properties of the function $\sigma(t)$. 

$\sigma(t)$ is an almost periodic function since the
sum of the moduli of its Fourier coefficients is finite. Namely,
by (\ref{eq:deltaeval}), the terms in the series (\ref{eq:sigma}) 
defining $\sigma(t)$ are 
majorized by those of a convergent series (whose sum is $C\pi^{-1} |||W|||^2$).  
Therefore, the series in (\ref{eq:sigma}) is uniformly convergent. As the uniform
limit of almost periodic functions, $\sigma(t)$ is then itself almost periodic,
bounded by 
\be
\sup_{t\in\R}|\sigma(t)|\ \le\ C|||W|||^2
\label{eq:sigmabound}
\ee
for some constant $C$; see also section \ref{se:app}. Moreover, $\sigma(t)$ has  mean value zero
since all the Fourier exponents are nonzero; see (\ref{eq:sigma}) 
 and section \ref{se:app}.
Therefore 
\be
\int_\tau^t\sigma(s)ds\le \frac{\Gamma}{2}(t-\tau),\  {\rm for}\ 
 \ t-\tau\ge {\cal M}\label{eq:sigmaest}
\ee
provided 
${\cal M}$ is taken sufficiently large.
It can be shown (see section \ref{se:app} or \cite{kn:Bohr}, page 42) that 
  (\ref{eq:sigmaest})
 holds provided   
\be
{\cal M}\ge \frac{4\ \sup_{t\in\R}\{|\sigma(t)|\}\ L(\Gamma/4)}{\Gamma/2}.
\label{eq:calMdef}
\ee
where $L(\Gamma/4)$ (see Definition \ref{de:apf}) 
 is such that in each interval of length $L(\Gamma/4)$ there
is at least one $\Gamma/4$-almost period
for $\sigma$.

Using (\ref{eq:sigmabound}) and then {\bf (H5)}, we can choose:
\be
{\cal M}=8CL(\Gamma/4)/ \theta_0 \label{eq:defcalM}
\ee
independently of $\ve$ and still satisfy (\ref{eq:calMdef}).

We now return to  the estimation of $R_A$. 
We split the integral in
(\ref{eq:prop4aa1}) into two integrals, one from $0$ to $t-{\cal M}$ and the
other from $t-{\cal M}$ to $t$. For the former we use (\ref{eq:sigmaest}) while
for the latter we use (\ref{eq:sigmabound}).  The result is
\ba
|R_A(t)|\ &\le&\ \int_0^{t-{\cal M}}\ e^{-{{1\over2}\ve^2\Gamma}(t-\tau)}|\tilde E(\tau)|d\tau 
 \no\\
 &+& \int_{t-{\cal M}}^t\ e^{\ve^2(C|||W|||^2-\Gamma)(t-\tau)}|\tilde E(\tau)|d\tau.
\label{eq:RAestimate11}
\ea
The first integral in (\ref{eq:RAestimate11}) can be bounded exactly as the 
term $\int_0^t\ e^{-\ve^2\Gamma (t-\tau)}|\tilde E(\tau)|d\tau$ in the proof of 
Proposition \ref{pr:odeest}. The second integral in (\ref{eq:RAestimate11}) is bounded in 
the following manner:
\ba
\lefteqn{\lan t\ran^{r_1}\int_{t-{\cal M}}^t\ e^{\ve^2(C|||W|||^2-\Gamma)(t-\tau)}|\tilde
E(\tau)|d\tau \le}\no\\
& & \frac{\lan t\ran^{r_1}}{\lan t-{\cal M}\ran^{r_1}}\int_{t-{\cal M}}^t\
e^{\ve^2(C|||W|||^2-\Gamma)(t-\tau)}d\tau\sup_{t-{\cal M}\le\tau\le t}
\left(\lan\tau\ran^{r_1}|E(\tau)|\right)\no\\
&\le& D\ \sup_{(\ve^2\Gamma/2)^{-\alpha}\le\tau\le t}
\left(\lan\tau\ran^{r_1}|E(\tau)|\right).\label{eq:RAestimate12a}
\ea
Note that $\ve$ and consequently $\ve^2\Gamma\sim \ve^2|||W|||^2$ are small, so we can
consider ${\cal M}\ll (\ve^2\Gamma/2)^{-\alpha}$ and $D\sim {\cal M}\ll (\ve^2\Gamma)^{-1}$.
 The result is (\ref{eq:prop4b1}).
A simple  bound, using the definition of $R_A(t)$ yields (\ref{eq:prop4c1}).

This completes the proof of Proposition \ref{pr:odeest1}.

\section{Dispersive Estimates and Local Decay.}\label{se:localdecay}

In this section we prove the local decay of $\phi_d$ and the decay in 
time of the remainder terms, $E(t)$,  in bound state amplitude equation
 (\ref{eq:Aeqn3}) of section \ref{se:normalform}. The arguments rely on 
hypotheses {\bf (H1)-(H5)} and  results of the previous section, so we
will handle Theorem \ref{th:twofive} first.
However, due to the differences between Theorems \ref{th:twosix} and
\ref{th:twotwo} we separately finish their proofs in the final two subsections of this section.  
 We will repeatedly use the following:

\begin{lem}\label{lm:sixone} For any $\eta\in [0,r_1]$ and $j\in\Z$ we have
\be
\left\| \int^t_0 \wmi e^{-iH_0(t-s)} \Pc f(s)ds \right\|\ \le\  
 C\lan t\ran^{-\eta} \sup_{0 \le\tau\le t} 
 \left( \lan\tau\ran^{\eta}\|\wpl f(\tau) \|\right)
 \label{eq:localdecayintegral}
\ee
and
\be
\left\| \int^t_0 \wmi e^{-iH_0(t-s)} \Pc (H_0-\la_0-\mu_j-i0)^{-1}f(s)\ ds 
 \right\|\  
 \le\ C\lan t\ran^{-\eta} \sup_{0\le\tau\le t}
 \left(\lan\tau\ran^{\eta}\left\|\wpl f(\tau)\right\|\right). 
  \label{eq:localdecaysingular}
\ee
\end{lem}

\noindent
\underline{Proof.}  The proof follows from the assumed local decay estimates 
on $e^{-iH_0t}$; see {\bf (H3a)}. Namely, using that $r_1>1$, 
\ba
\left\| \int^t_0 \wmi e^{-iH_0(t-s)} \Pc f(s)\ ds \right\|\  
&\le&\  \int_0^t \| \wmi e^{-iH_0(t-s)}\ \Pc \wmi \|_{{\cal L}({\cal H})}
 \lan s\ran^{-\eta}\ ds\no
\\
&&\ \cdot\sup_{0\le\tau\le t} \left(\lan\tau\ran^{\eta} 
	 \|\wpl f(\tau )\| \right)\no 
  \\
  &\le&\ C\int_0^t \lan t-s \ran^{-r_1}\lan s\ran^{-\eta}\ ds\  
\sup_{0 \le\tau\le t}\left( \lan\tau\ran^{\eta} \|\wpl f(\tau )\| \right)\no
 \\
 &\le& C\lan t\ran^{-\eta} \sup_{0 \le\tau\le t}\left(\lan\tau\ran^{\eta}
  \left\| \wpl f(\tau ) \right\|\right)\no 
\ea
which proves (\ref{eq:localdecayintegral}).  
 The proof of (\ref{eq:localdecaysingular}) is identical, and uses the
singular local decay estimate of {\bf (H3) (b)} \ \ \ \ \  []. 

We now define the norms 
\be
 [A]_\alpha(T)\ =\ \sup_{0\le\tau\le T} 
  \lan\tau\ran^\alpha |A(\tau)| \nonumber
 \ee
 and 
 \be
 [\phi_d]_{LD,\alpha}(T)\ =\ \sup_{0\le\tau\le T} \lan\tau\ran^\alpha 
 \| \wmi\phi_d(\tau) \|\no
\ee
Then we have

\begin{prop}\label{pr:phidestimate}
For any $T>0$ and $\eta\in [0,\ r_1],$  
\begin{equation}  
[\phi_d]_{LD,\eta}(T) \le C 
  \left(\ \|\wpl\phi_d(0)\|\ +\ |\ve|\  
  ||| W |||\  [A]_{\eta}(T)\ \right) .
\label{eq:phidestimate}
\end{equation}
\end{prop}

\noindent
\underline{Proof.}
 From equation (\ref{eq:intphideqn}) we get, using the assumed 
local decay estimate for $e^{-iH_0t}$ and (\ref{eq:localdecayintegral}),
 
\ba
\left\| \wmi\phi_d(t) \right\|\ &\le &\  \sum_{j=0}^2\ \|\wmi\phi_j(t)\|\no\\
 &\le &\ C\lan t\ran ^{-\eta} 
 \left\| \wpl \phi_d(0) \right\|
+\  C|\ve|\lan t\ran ^{-\eta}\  [A]_{\eta}(t)\ 
 \sup_{0\le s\le t}\|\wpl W(s)\psi_0\|\no\\
&+&\  C\ |\ve|\ ||| W |||\ \lan t\ran^{-\eta}\left[\phi_d\right]_{LD,\eta}(t).
\label{eq:phixyz}
\ea

Since $\|\wpl W(s)\psi_0\|\le ||| W |||\ \|\psi_0\|\ =\ |||W|||$ 
 and $|\ve|\ |||W|||$ is 
 assumed to be small, multiplying both sides of this last equation 
  by $\lan t\ran^{\eta}$ and taking supremum over $t\le T$ yields
  (\ref{eq:phidestimate}).
  \ \ \ \ \ []

We now estimate $E(t)$.

\begin{prop}\label{pr:Eestimate} Let $T>0$. For any $\eta\in [0,r_1]$: 
\be
  [E]_{\eta}(T)\  
\le\ C \left(\ \ve^2|||W|||^2\  |A(0)|\  +\  |\ve|\ |||W|||\  \|\wpl\phi_d(0)\|\
+\ |\ve|^3|||W|||^3\  [A]_{\eta}(T)\ \right)    
\label{eq:Eestimate}
\ee
\end{prop}

\noindent
\underline{Proof.}

$E(t)$ is defined in (\ref{eq:Eexpand}). From these
equations it is seen that we need to bound
 the following terms:
\ba
R_1 &\equiv & {1\over4}\ve^2
  |A(0)|\sum_{j,k\in\Z}\left|
 \left(\beta_k\psi_0,\ e^{-iH_0t}\ (H_0 - \la_0 -\mu_j-i0)^{-1} 
 \Pc \beta_j\psi_0\right)\right|\no  \\ 
R_2 &\equiv &  {1\over 4}\ve^2\sum_{j,k\in\Z}\left|  
 \left(\beta_k\psi_0, \int_0^t\ e^{-iH_0(t-s)}(H_0 - \la_0 -\mu_j-i0)^{-1}
  \Pc e^{-i(\la_0+\mu_j)s}\D_sA(s)\beta_j\psi_0 ds\right)\right|\no
\ea
and
\ba
\left|\ve\left(\psi_0,\  W(t)\phi_0 (t)\right)\right| 
 &=& \left|\ve\left(W(t)\psi_0,e^{-iH_0t}\phi_d (0)\right)\right|\nonumber \\
\left|\ve\left(\psi_0,\  W(t)\phi_2\right)\right|
 &=& \left|\ve^2\left(W(t)\psi_0,\int_0^t e^{-iH_0(t-s)}\Pc W(s)\phi_d (s)ds\right)
 \right|.\no
\ea
  The estimates of the above terms repeatedly use Lemma \ref{lm:sixone}. 
  Let $\eta\in [0,r_1]$. 

\nit
\un{Estimation of $R_1$:}

\ba
R_1 &=& {1\over4}\ve^2|A(0)|\sum_{j,k\in\Z}\left|
\left(\wpl\beta_k\psi_0,\  \wmi e^{-iH_0t}(H_0-\la_0-\mu_j-i0)^{-1}\Pc\wmi\ 
	\wpl\beta_j\psi_0\right)\right|\no\\
&\le & C |A(0)|\ \ve^2|||W|||^2\ \lan t\ran^{-\eta}
\label{eq:alpha1est}
\ea
by the local decay estimates (\ref{eq:singularldest}).

\medskip
\noindent
\un{Estimation of $R_2$}

From (\ref{eq:Aeqn3}) we have that
\be
|\D_sA(s)| \le C |\ve|\ |||W||| \ |A(s)| + |E(s)|  \label{eq:DAofsestimate} 
\ee
since $\Im\rho$ is linear in $|\ve|\ |||W|||$ and $\Re\rho$, $\Gamma$ are quadratic.

Applying Lemma \ref{lm:sixone} to $R_2$ we then get
\ba
R_2\ &=&{1\over 4}\ve^2\sum_{j,k\in\Z}\left|  
 \left(\wpl\beta_k\psi_0, \int_0^t\wmi e^{-iH_0(t-s)}(H_0 - \la_0 -\mu_j-i0)
 ^{-1}\Pc\wmi\D_sA(s)\wpl\beta_j\psi_0 ds\right)\right|\no\\
&\le&\ C \ve^2 ||| W|||^2\ \lan t\ran^{-\eta} 
 \left( |\ve|\  |||W|||\ [A]_{\eta}(t)\ +\ [E]_{\eta}(t)\right).
 \ea

\medskip
\noindent
\underline{Estimation of  $\left|\ve\left(\psi_0,\  W(t)\phi_0 (t)\right)\right| $:}

Since, by definition, $\phi_d(0) = \Pc\phi_d(0)$ 
 we can apply local decay estimates for $e^{-iH_0 t}$  to get
\be
\left|\ve\left(W(t)\psi_0,\  \phi_0 (t)\right)\right|  \le\  
	 C|\ve|\ |||W|||\ \lan t\ran^{-\eta} \; 
 \|\wpl\phi_d(0)\|.   
 \label{eq:alpha5est}
 \ee

\medskip
\noindent
\underline{Estimation of $\left|\ve \left(\psi_0,\  W(t)\phi_2\right)\right|$}

Applying Lemma \ref{lm:sixone} as before we get, for $0\le t\le T$, 
\be
\left|\ve\left(\psi_0,\  W(t)\phi_2\right)\right| \le C\ve^2|||W|||^2\  
 \lan t\ran^{-\eta} \; [\phi_d]_{LD,\eta}(T). \label{eq:temp}
\ee
Using Proposition \ref{pr:phidestimate} to estimate $[\phi_d]_{LD,\eta}(t)$ 
in (\ref{eq:temp}), we get
\be
\left|\ve\left(\psi_0,\  W(t)\phi_2\right)\right|
 \le C\ve^2|||W|||^2\ \lan t\ran^{-\eta} \left\{\ \|\wpl\phi_d(0)\|\ +\ 
 |\ve|\ |||W|||\  [A]_{\eta}(t)\ \right\}.  
\no\ee

Finally, combining the above estimates, 
 we can  
 bound $[E]_{\eta}(T)$ for any $\eta\in [0,r_1]$  as follows: 

\be
 [E]_{\eta}(T) \le C \left\{ 
 \ve^2|||W|||^2\ |A(0)| + |\ve|\ |||W|||\ \|\wpl\phi_d(0)\| + \ve^2|||W|||^2\  [E]_{\eta}(T)
  + |\ve|^3|||W|||^3\ [A]_{\eta}(T)\right\} .
\ee Since $|\ve|\ |||W|||$ is assumed to be small, Proposition \ref{pr:Eestimate}  follows.
 \ \ \ \ [] 
\medskip

We can now complete the proof of 
 Theorem \ref{th:twofive}.  To prove the assertions concerning the
 infinite time behavior, the key is to establish local decay
 of $\phi_d$, in particular, the uniform boundedness of $[\phi_d]_{LD,r_1}(T)$.
 This will follow directly from Proposition \ref{pr:phidestimate} 
 if we prove the uniform boundedness 
 $[A]_{r_1}(T)$, or equivalently $[\tilde{A}]_{r_1}(T)$.

\begin{prop}\label{pr:Aestimate} Under the hypothesis of Theorem \ref{th:twofive},
 there exists an $\ve_0>0$ such that for each real number $\ve,\ |\ve|<\ve_0$ there is a constant $C_*$,
  with the property that for any  $T>0$
$$
[A]_{r_1}(T) \le C_*
$$
\end{prop}

\noindent
\underline{Proof.}

We begin with the expansion of $A(t)$ given in Proposition \ref{pr:odeest1}.
Multiplying (\ref{eq:prop4a1}) by $\lan t\ran^{r_1}$, and taking the
supremum over $0\le t\le T$ we have:
\be
[A]_{r_1}(T)\ \le\  
 C\left( |A(0)|\ (\ve^2\Gamma/2)^{-r_1}\ +\ 
 \sup_{0\le \tau\le 2(\ve^2\Gamma/2)^{-\alpha}} \lan\tau\ran^{r_1}|R_A(\tau)|\ +\ 
  \sup_{2(\ve^2\Gamma/2)^{-\alpha}\le \tau\le T} \lan\tau\ran^{r_1}|R_A(\tau)|\ \right)
 \label{eq:oops}\ee
The right hand side of (\ref{eq:oops}) is estimated using Proposition
\ref{pr:odeest1}.
\ba
[A]_{r_1}(T)&\le & 
  C|A(0)|\ (\ve^2\Gamma/2)^{-r_1}\ + 
  D\ (\ve^2\Gamma/2)^{-\alpha (r_1-1)}\ [E]_0(2(\ve^2\Gamma/2)^{-\alpha})\no\\
&+&
  C_1\ e^{-(\ve^2\Gamma/2)^{-\delta}}\ [E]_0(2\MM)\ + C_2\ (\ve^2\Gamma/2)^{-1}[E]_{r_1}(T).
 \no\ea
 Next, we apply Proposition \ref{pr:Eestimate} which yields:
 \ba
[A]_{r_1}(T)\ &\le&\  
 C |A(0)|\ (\ve^2\Gamma/2)^{-r_1}\ +\  
D\ (\ve^2\Gamma/2)^{-\alpha (r_1+1)}\ [E]_0(2\MM)\no\\
&+& 
 C_1 e^{-(\ve^2\Gamma/2)^{-\delta}}\ [E]_0(2\MM)\label{eq:last-est}\\
 &+&\ C_2(\ve^2\Gamma/2)^{-1}\left( \ve^2|A(0)| |||W|||^2\ +\ |\ve|\ 
 |||W|||\ \|\wpl\phi_d(0)\|\ +\ |\ve|^3|||W|||^3\ [A]_{r_1}(T)\right).\no
\ea

 Note that by Proposition \ref{pr:Eestimate} and the simple bound: 
 $$
 [A]_0(T)\ \le \ 
   \|\phi_0\|, 
  $$
  $[E]_0(2\MM)$ 
 is  bounded in terms of the initial data and $|\ve|\ |||W|||$.

Choose $\ve_0$ such that:
$$1-\frac{2C_2|||W|||^3}{\Gamma}\ve_0=0,$$ where $C_2$ is the same as in (\ref{eq:last-est}). 
Then, for $|\ve|<\ve_0$
\be
[A]_{r_1}(T) \le C_* 
\label{eq:this-est}
\ee 
 Here, $C_*$  depends on 
$\|\phi_0\|,\ \|w_+\phi_0\|,\ r_1
 $ and  $\ve$.

This completes the proof of Proposition \ref{pr:Aestimate} and therewith the $t\to\infty$
asymptotics asserted in Theorems \ref{th:twofive}-\ref{th:twotwo}. \ \ \ \ [] \bigskip

%

It remains to finish the proofs of the Theorems \ref{th:twotwo} and
\ref{th:twosix}. Due to some differences we consider them separately in
the following two subsections. 

\subsection{Proof of Theorem \ref{th:twotwo}}

In order to obtain (\ref{eq:projection}) we note that (\ref{eq:Adef}),
(\ref{eq:prop4a}) and (\ref{eq:rararel1}) together with the definition of $\omega(t)$ in (\ref{eq:omegadef}) already gives us:
\ba
a(t)&=&e^{-i\la_0 t +\int_0^t\rho(s)ds}\left(A(0)e^{-\ve^2\Gamma t}+R_A(t)\right)\no\\
&=&a(0)e^{-\ve^2\Gamma t}e^{i\omega(t)}+R_a(t).\no
\ea
which is in fact the second relation in (\ref{eq:projection}). The third is a direct consequence of the second since $P(t)=|a(t)|^2$ while
the fourth relation is exactly (\ref{eq:intphideqn}).
 
It remains to prove the intermediate time estimate (\ref{eq:Raestimate}).
 The ingredients are contained in (\ref{eq:Eestimate}) and its proof.
 First, by (\ref{eq:rararel12})
 $$
 |R_a(t)|\ \le C\ |R_A(t)|\ +\ {\cal O}(\ve^2|||W|||^2) 
 $$
So, it suffices to prove an ${\cal O}(|\ve|\ |||W|||)$ upper bound for $R_A$.

Using (\ref{eq:prop4aa}) we know that
\be
|R_A(t)|\le \int_0^t\ e^{-\ve^2\Gamma(t-\tau)}\ \left| E(\tau)\right|\ d\tau .\label{eq:RA}
\ee 
Let $T_0$ denote an arbitrary  fixed positive  number. We estimate
 (\ref{eq:RA}) for $t\in [0, T_0(\ve^2\Gamma)^{-1}]$. We bound the exponential in
 the integrand by one (explicit integration would give something of order
  $(\ve^2\Gamma)^{-1}$), and bound $|E(\tau)|$ by estimating the expressions 
  in the proof of Proposition \ref{pr:Eestimate}. First, the
  estimates of Proposition \ref{pr:Eestimate} for $R_1$ and 
  $\left|\ve\left(\psi_0,\  W(t)\phi_0 (t)\right)\right|$ are
  useful as is. Integration of the bounds (\ref{eq:alpha1est})
  and (\ref{eq:alpha5est}) gives:
  \ba
  \int_0^t\ e^{-\ve^2\Gamma (t-\tau)}R_1\ d\tau &\le & 
   C\ \ve^2|||W|||^2\ \|w_+\phi(0)\|, \no\\
  \int_0^t\ e^{-\ve^2\Gamma (t-\tau)} 
  \left|\ve\left(\psi_0,\  W(t)\phi_0 (t)\right)\right|\ d\tau &\le & 
   C\ |\ve|\ |||W|||\ \|w_+\phi(0)\|. 
   \label{eq:est135}
   \ea
   To estimate the contributions of $R_2$, first
   observe that by (\ref{eq:DAofsestimate}) and Proposition \ref{pr:Eestimate}
   with $\eta=0$ 
   \be
   |\D_sA(s)|\ \le\ C\ |\ve|\ |||W|||\ \|w_+\phi(0)\| 
   \nn\ee
   Therefore, using local decay estimates we have:
   \ba
   \int_0^t\ e^{-\ve^2\Gamma (t-\tau)}R_2\ d\tau&\le&\ 
	C\ T_0(\ve^2\Gamma)^{-1}\ |\ve|^3|||W|||^3\  \|w_+\phi(0)\|
	\nn\\
	&\le&\ D|\ve|\ |||W|||\ \|w_+\phi(0)\|.
	\nn\ea
Finally, we come to the contribution of 
$\left|\ve\left(\psi_0,\  W(t)\phi_2\right)\right|$. We rewrite it as follows.
\ba
\left|\ve\left(\psi_0,\  W(t)\phi_2\right)\right| &=&\ \ve^2 
 \left|\int_0^t\left(W(s)e^{iH_0(t-s)}\Pc W(t)\psi_0, \phi_d
 (s)\right)ds\right|\no\\ 
&=&\ \left| \int^t_0 \ve^2( w_+W(s) w_+\cdot w_- e^{iH_0(t-s)}\Pc w_-\cdot
  w_+W(t)\psi_0,\
   w_-\phi_d(s))\ ds \right|.\label{eq:alpha6rewrite}\ea
Recall that by (\ref{eq:intphideqn}) 
 $\phi_d\ =\ \phi_0\ +\ \phi_1\ +\ \phi_2$,  
  where $\phi_0(t)=e^{-iH_0t}\phi_d(0)$. Using local decay estimates {\bf
  (H3a)}, the
  contribution of the term $\phi_0(t)$ can be bounded by 
  $C\ \ve^2|||W|||^2\ \|w_+\phi_d(0)\|\ \lan \tau\ran^{-r_1}$. Multiplication of this
  bound by $e^{-\ve^2\Gamma(t-\tau)}$ and integration with respect to $t$
  gives the bound $C\ \ve^2|||W|||^2 \|w_+\phi_d(0)\|$.
   To assess the contributions from $\phi_1+\phi_2$,
  note that local decay estimates {\bf (H3a)} imply  
  \be 
  \|w_-(\phi_1+\phi_2)\|\ \le\ C\ |\ve|\ |||W|||\ \| w_+\phi(0)\|.
  \nn\ee
  Putting together the contributions from $\phi_0$ and from
  $\phi_1+\phi_2$, we have:
  \be
  \int_0^t\ e^{-\ve^2\Gamma (t-\tau)}\left|\ve\left(\psi_0,\  W(t)\phi_2\right)\right| \ d\tau
  \ \le\ C\left(\ \ve^2|||W|||^2\ \|w_+\phi_d(0)\|\ +\ 
  (\ve^2\Gamma)^{-1}\ |\ve|^3|||W|||^3\ \right)
   \nn\ee
The above estimates and (\ref{eq:rararel12}) imply 
  (\ref{eq:Raestimate}). Now, (\ref{eq:Raprimestimate}) is a direct consequence of (\ref{eq:Raestimate}) and the relation $P(t)=|a(t)|^2$.
This concludes the proof of Theorem \ref{th:twotwo}.

\subsection{Proof of Theorem \ref{th:twosix}}

As in the proof of Theorem \ref{th:twotwo} relations (\ref{eq:Adef}),
(\ref{eq:prop4a1}), (\ref{eq:rararelation}) and the definition of $\omega(t)$ in (\ref{eq:omegadef}) gives:
\ba
a(t)&=& e^{-i\la_0t+\int_0^t\eta(s)ds}\left(A(0)e^{-\ve^2(\Gamma
t-\int_0^t\sigma(s)ds)}+R_A(t)\right)\no\\
&=&a(0)e^{-\ve^2(\Gamma-\gamma)t}e^{i\omega(t)}+R_a(t),\no
\ea
which is the second relation in (\ref{eq:projection1}). In what follows, the only difference from the previous argument is in estimating $R_a(t)$.

We start with the relation (\ref{eq:rararelation}):
\be
R_a(t)\ =\ e^{-i\la_0t+\int_0^t\eta(s)ds}R_A(t)+
\left(1-e^{\ve^2(\int_0^t\sigma(s)ds-\gamma t)+\Re\int_0^t\eta(s)ds}\right)e^{-\ve^2(\Gamma-\gamma )t}a(0).
\label{eq:rara1}\ee
Since $\sigma(t)$ is an almost periodic function with zero mean, for any $\gamma>0$ there is an ${\cal M}_{\gamma}>0$ such that whenever
$|t|\ge {\cal M}_{\gamma}$
$$\int_0^t\sigma(s)ds\le \gamma t.$$
On the
other hand for $|t|<{\cal M}_{\gamma}$, using (\ref{eq:sigmabound}) we have:
$$\int_0^t\sigma(s)ds\le C{\cal M}_{\gamma}|||W|||^2.$$ So, in both cases,
$$\int_0^t\sigma(s)ds-\gamma t\le C{\cal M}_{\gamma}|||W|||^2.$$
Substituting now in (\ref{eq:rara1}) and tacking into account that by (\ref{eq:varrhobounds1}),
$$ \Re\ \int_0^t \eta(s)ds\ \le\ C\ve^2|||W|||^2 $$ uniformly in $t$, we get
\be
\left|R_a(t)\right|\ \le\ C\left|R_A(t)\right|+{\cal O}(\ve^2|||W|||^2)
\label{eq:rara2}\ee

It remains to prove an ${\cal O}(|\ve|\ |||W|||)$ for $R_A(t)$.
 Looking now at (\ref{eq:RAestimate11}) we
see that we can bound the exponential by $\max\{1,e^{\ve^2(C|||W|||^2-\Gamma){\cal M}}\}$.
Now, the same argument as in the end of the previous subsection will give us
the required result. 

This completes the proof of Theorem \ref{th:twosix}.

\section{Generalizations}\label{se:generalizations}

In the previous sections we considered perturbations of the form $\ve W(t)$, 
with $W(t)$ independent of $\ve$. In this section, we shall extend our theory 
to a more general class of potentials, $W_{\ve}$, which are small for small 
$\ve$, but which may deform nontrivially as $\ve$ varies.

Consider a family of perturbations ${\cal W}$ and the general system
\ba
i \D_t\phi(t) & = & \left(H_0 + W(t))\right)\phi(t) , \no \\
\phi|_{t=0} & = & \phi(0). \label{eq:newgeneralse}
\ea
where $W\in{\cal W}$ (compare to (\ref{eq:generalse}). The results are

\begin{theo}\label{th:twothree}
 Suppose that $H_0$ and any $W\in{\cal W}$ satisfy hypotheses {\bf (H1)-(H5)}.
 In addition assume:
\medskip

\nit {\bf (H7)} \underline{Equi-almost periodicity}: 
There exists a positive constant $L_{\theta_0}$, independent of $W\in {\cal W}$, 
such that in any interval of real numbers of length 
 $L_{\theta_0}$,
the  function $|||W|||^{-2}\ \sigma(t)$\  ($|||W|||\ne0$), where
\be
\sigma(t)\equiv -\frac{\pi}{4}\Re \sum_{j\in\res ,j\neq k\in\Z}e^{i(\mu_k-\mu_j)t}
\left(\ \beta_k\psi_0,\delta(H_0-\la_0-\mu_j)\beta_j\psi_0\right).
 \label{eq:newsigma}
\ee
 has 
 a $\theta_0/4$ almost period, $\theta_0$ is given by {\bf (H5)}. More precisely,
  there exists $L_{\theta_0}>0$ which 
does not depend on $W$ such that in any interval of length $L_{\theta_0}$ there 
is a number $\tau=\tau(\theta_0/4)$, such that for all $t\in\R$
\be
\left|\ |||W|||^{-2}\sigma(t+\tau)\ -\ |||W|||^{-2}\sigma(t)\ \right|\ \le\
 \theta_0/4.\no\ee

If $w_+\phi(0)\in{\cal H}$, 
then, there exists an $\ve_0 >0$ (depending on ${\cal C},\ r_1,\ \theta_0$ and $L_{\theta_0}$)
such that whenever $||| W |||< \ve_0$, 
 the solution of (\ref{eq:newgeneralse}) satisfies the local decay
estimate (identical with the one in Theorem \ref{th:twofive}):
\begin{equation}
\| \wmi\ \phi(t)\| \le C\lan t\ran^{-r_1} \| \wpl\ \phi_0\|,\ \ t\in\R.
\label{eq:newldestimate}
\end{equation} 
\end{theo}

\nit\un{Sketch of Proof.} Once we drop $\ve$ from all expressions (since it is 
not present in the actual setting), the arguments in the previous sections hold
 in this case except the analysis of $\sigma(t)$ in Proposition 
\ref{pr:odeest1}. Formulas (\ref{eq:sigma}) and (\ref{eq:newsigma}) are the 
same, but now $\mu_j,\ \beta_j,\ j\in\Z$ are not fixed as they define $W$ by 
{\bf (H4)} and $W$ sweeps a general class ${\cal W}$. This may prevent us to 
find a fixed time interval, ${\cal M}$, independent of $W\in{\cal W}$, after 
which $\sigma(t)$ is within $\Gamma/2$ distance from its mean, see relations 
(\ref{eq:sigmaest}-\ref{eq:defcalM}).

Nevertheless, {\bf (H7)} is exactly what we need to overcome the difficulty. 
A straightforward calculation shows that any $\theta_0/4$ almost period of 
$|||W|||^{-2}\sigma(t)$ is a $\Gamma/4$ almost period for $\sigma(t)$. 
Consequently, $L(\Gamma/4)$ in (\ref{eq:defcalM}) is bounded above by 
$L(\theta_0)$ given in {\bf (H7)}. But the latter is fixed, so, we can choose
\be
{\cal M}=8CL(\theta_0)/\theta_0 \label{eq:newdefcalM}
\ee
independent of $W\in{\cal W}$ and still satisfy (\ref{eq:calMdef}) hence (\ref{eq:sigmaest}).

Finally, we can close the arguments exactly as we did for Theorem \ref{th:twofive}.  
\medskip

\begin{rmk}
Theorems analogous to Theorem \ref{th:twosix} (respectively Theorem 
\ref{th:twotwo}) can be proved under hypotheses {\bf (H1)-(H5), (H7)} 
(respectively {\bf (H1)-(H6)}).
\end{rmk}

\nit {\bf Examples:} {\bf (H7)} holds trivially for
 
\nit {\bf (1)} $ {\cal W}=\{\ve W(t,x):\ \ve\in\R,\ W\ {\rm fixed}\}$

\nit or

\nit {\bf (2)} ${\cal W}=\{\ve W(\ve^{-1}t,x):\ \ve\in\R-\{0\},\ |\ve|\le 1,\ \ W\ {\rm fixed}\}.$
\medskip

In the Example 1, $\ve$ cancels in the formula $|||W|||^{-2}\sigma(t)$ while 
in the Example 2 we have a time dilatation which shrinks the gaps between the 
almost periods, so $L(\theta_0)$ valid for $W$ is good for the entire family.
\medskip 

\nit {\bf (3)} There are more general families of perturbations ${\cal W}$ for which 
{\bf (H7)} holds. For example if ${\cal W}$ is equi-almost periodic, see section \ref{se:app}.

%
\section{Appendix: Singular operators}\label{se:sop}

In this section we present the definition and the properties we needed previously for the singular operators:
$$ e^{-iH_0t}\left(H_0-\Lambda-i0\right)^{-1}\Pc ,\ \delta\left(H_0-\Lambda\right)\Pc,\ {\rm P.V.}\left(H_0-\Lambda\right)^{-1}\Pc,
$$
and establish the identities
$$
\left(H_0-\Lambda\mp i0\right)^{-1}\Pc\ =\ {\rm P.V.}\left(H_0-\Lambda\right)^{-1}\Pc \pm i\pi\delta\left(H_0-\Lambda\right)\Pc 
$$
suggested by the well known distributional identities
$$
(x\mp i0)^{-1}\ =\ {\rm P.V.}\ {1\over x}\ \pm i\pi\ \delta(x).
$$

Recall that we are in the complex Hilbert space ${\cal H}$ with self-adjoint ``weights" $w_\pm$ and projection operator $\Pc$ satisfying (i), (ii) and (iii). We can then construct the complex Hilbert space ${\cal H_+}$ as the closure of the domain of $\wpl$ under the scalar product $\left(f,g\right)_+=\left(\wpl f,\wpl g\right)$ and the complex Hilbert space ${\cal H_-}$ as the closure of ${\Pc\cal H}$ under the scalar product $\left(f,g\right)_-=\left(\wmi f,\wmi g\right)$ . 

By hypotheses of section \ref{se:mainresults}, $H_0$ is a self-adjoint operator on ${\cal H}$ and satisfies the local decay estimate (\ref{eq:localdecay}). Based on this property, in \cite{kn:MS,kn:nh,kn:TDRT1} it is proved that for $\Lambda$ in the continuous spectrum of $H_0$ and $t\in\R$
\ba
T_t&\equiv &i\lim_{\eta\searrow 0}\int_t^\infty e^{-i\left(H_0-\Lambda-i\eta\right)s}ds\Pc \no\\
T_t^*&\equiv &-i\lim_{\eta\searrow 0}\int_{-\infty}^{-t} e^{-i\left(H_0-\Lambda+i\eta\right)s}ds\Pc \no
\ea
are well defined, linear bounded operator from ${\cal H_+}$ to ${\cal H_-}$. We then define
\ba
 e^{-iH_0t}\left(H_0-\Lambda-i0\right)^{-1}\Pc &\equiv & e^{-i\Lambda t} T_t \label{eq:1sop}\\
 e^{+iH_0t}\left(H_0-\Lambda+i0\right)^{-1}\Pc &\equiv & e^{+i\Lambda t} T_t^*, \label{eq:2sop}
\ea
and 
\ba
{\rm P.V.}\left(H_0-\Lambda\right)^{-1}\Pc &\equiv &{1\over 2}(T_0+T_0^*) \label{eq:PVdef}\\
\delta\left(H_0-\Lambda\right)\Pc &\equiv &{1\over2\pi i}(T_0-T_0^*) \label{eq:deltadef}
\ea
Note that the definitions imply the identities:
\be\label{eq:distrid}
 \left(H_0-\Lambda\mp i0\right)^{-1}\Pc = {\rm P.V.}\left(H_0-\Lambda\right)^{-1}\Pc \pm i\pi\delta\left(H_0-\Lambda\right)\Pc.  
\ee
Particularly important properties of these operators are their symmetries when viewed as quadratic forms on ${\cal H_+}\times {\cal H_+}$. For example, on any $f,\ g\in{\cal H_+}$ the quadratic form induced by $T_t$ is given by:
$$
\left(f, g\right)\mapsto (\wpl f, \wmi T_t g).
$$
Note that
\be\label{eq:quadid}
\lim_{\eta\searrow 0}\left(f,T_t^\eta g\right)\equiv\lim_{\eta\searrow 0}\left(f,i\int_t^\infty e^{-i\left(H_0-\Lambda-i\eta\right)s}ds\Pc g\right)\ =\ (\wpl f, \wmi T_t g)
\ee
by the following calculation:
\ba
\lim_{\eta\searrow 0}\left(f,i\int_t^\infty e^{-i\left(H_0-\Lambda-i\eta\right)s}ds\Pc g\right) &=&\lim_{\eta\searrow 0}\left(f,\Pc i\int_t^\infty e^{-i\left(H_0-\Lambda-i\eta\right)s}ds\Pc g\right) \no\\
&=&\lim_{\eta\searrow 0}\left( f,\wpl\wmi\Pc i\int_t^\infty e^{-i\left(H_0-\Lambda-i\eta\right)s}ds\Pc g\right) \no\\
&=&\lim_{\eta\searrow 0}\left(\wpl f,\wmi i\int_t^\infty e^{-i\left(H_0-\Lambda-i\eta\right)s}ds\Pc g\right) \no\\
&=&(\wpl f, \wmi T_t g),\no
\ea
where we used the that $\Pc$ is a projection operator commuting with the integral operator, the identity $\wpl\wmi\Pc =\Pc$ on ${\cal H}$,  the self adjointness of $w_\pm$ and $\Pc $, and $\lim_{\eta\searrow 0}\wmi T_t^\eta=\wmi T_t$ in ${\cal L(H_+,H)}$. 

Identity (\ref{eq:quadid}) suggests the notation:
$$
\left(f, g\right)\mapsto ( f, T_t g)
$$
for the quadratic form induced by $T_t$, where $(\cdot,\cdot)$ can formally be treated as the scalar product in ${\cal H}$. Moreover, (\ref{eq:quadid}) implies
$$
(f, T_t g)=(T_t^*f,g).
$$
Therefore, the quadratic form induced by ${\rm P.V.}(H_0-\Lambda)^{-1}\Pc$ is the symmetric part of the one induced by $T_0$ while $\delta(H_0-\lambda)\Pc$ induces the skew-symmetric part of it divided by the factor $i\pi$. As a consequence both the forms corresponding to the last two operators are symmetric.

In conclusion, for any $f,g\in Domain(\wpl),\ t\in\R$ and $\Lambda\in\spectrum $ we have:
\ba 
(f,e^{\mp iH_0t}(H_0-\Lambda\mp i0)^{-1}\Pc g)&\equiv& \left(\wpl f,\wmi e^{\mp iH_0t}(H_0-\Lambda\mp i0)^{-1}\Pc g\right)\no\\
&\le &C_t\ \|\wpl f\|\ \|\wpl g\| \label{eq:rezbound}\\ 
(f,\delta(H_0-\Lambda)\Pc g)&\equiv& \left(\wpl f,\wmi\delta (H_0-\Lambda)\Pc g\right)
\le \ {C_0\over\pi}\ \|\wpl f\|\ \|\wpl g\| \label{eq:deltabound}\\ 
(f,{\rm P.V.}(H_0-\Lambda)^{-1}\Pc g)&\equiv& \left(f,\wmi {\rm P.V.}(H_0-\Lambda)^{-1}\Pc g\right)
\le \ C_0\ \|\wpl f\|\ \|\wpl g\|. \label{eq:PVbound}
\ea
The inequalities are due to the boundedness of $T_t$, where $C_t$ denotes the norm of $T_t$ in ${\cal L(H_+,H_-)}$. Moreover, the following symmetry properties hold:
\ba 
(f,e^{\mp iH_0t}(H_0-\Lambda\mp i0)^{-1}\Pc g)&=&(e^{\pm iH_0t}(H_0-\Lambda\pm i0)^{-1}\Pc f,g)\no\\
(f,\delta(H_0-\Lambda)\Pc g)&=&(\delta(H_0-\Lambda)\Pc f,g)\no\\
(f,{\rm P.V.}(H_0-\Lambda)^{-1}\Pc g)&=&({\rm P.V.}(H_0-\Lambda)^{-1}\Pc f,g).\no
\ea

\section{Appendix: Almost periodic functions}\label{se:app}

In this section we present the definition and the properties of almost periodic
functions we used throughout this paper. We will confine to functions of the 
form
$f:\R\rightarrow X$ where $X$ is a complex Banach space with norm denoted by
$\|\cdot\|$.

\begin{defin}\label{de:apf}
We say that $$f:\R\rightarrow X$$ is \un{almost periodic} if and only if it 
is continuous and
for each $\ve>0$ there exists a length $L(\ve,f)>0$ such that in any closed 
interval of length greater or equal than $L(\ve,f)$ there is at least one 
$\tau$ with the property that for all $t\in\R$ we have
\be
\|f(t+\tau)-f(t)\|\le\ve. \label{eq:eap}
\ee
The number $\tau$ with the property above is called an $\ve$-almost period for
$f$.
\medskip

We say that the family, ${\cal F}$ of almost periodic functions is 
\un{equi-almost periodic} if $L(\ve, f)$ can be choosen independently of 
$f\in {\cal F}$.
\end{defin}

{\bf Example 1} Any continuous periodic function is almost periodic since for 
any $\ve>0$ we can choose the length $L(\ve)$ to be the period of the function.

\begin{theo}\label{th:bound}
Any almost periodic function has a relative compact image.
\end{theo}

The proof of the theorem can be found in \cite[Property 1, pp. 2]{kn:Levitan}. In
particular, any almost periodic function $f:\R\rightarrow X$ is in the Banach
space of all bounded and continuous functions on $\R$ with values in $X$,
$C(X)$, endowed with the uniform norm. The next result is Bochner's 
characterization of almost periodic functions, see for example
\cite[Bochner's theorem, pp. 4]{kn:Levitan}.

\begin{theo}(Bochner)\label{th:bochner}
Let $f:\R\rightarrow X$ be a continuous function. For $f$ to be almost periodic
it is necessary and sufficient that the family of functions $\{f(t+h)\},\
-\infty<h<\infty$ is relatively compact in $C(X)$.
\end{theo}

As a consequence of Bochner's criterion and Property 4 from 
\cite[pp. 3]{kn:Levitan} we have:

\begin{theo}
Suppose $X_1,\ X_2,\ldots ,X_{k+1}$ are Banach spaces, $f_i:\R\rightarrow X_i,\
1\le i\le k$ are almost periodic functions and $g:\prod_{i=1}^k\rightarrow
X_{k+1}$ is continuous. Than $g(f_1(t),\ f_2(t),\ldots,f_k(t))$ is an almost
periodic function.
\end{theo}

The last theorem has very important consequences in the theory of almost
periodic functions. We will list only those which are useful in our
presentation.

\begin{cor}\label{co:sum}
A finite sum of almost periodic functions with values in the same Banach space
is an almost periodic function.
\end{cor}

\begin{cor}\label{co:prod}
A product between a complex valued almost periodic function and an arbitrary
almost periodic function is an almost periodic function.
\end{cor}

\begin{cor}\label{co:linop}
If ${\cal H}$ is a complex Hilbert space, ${\cal L}({\cal H})$ is the Banach
space of the bounded linear operators on ${\cal H}$ and 
$W:\R\rightarrow {\cal L}({\cal H})$ is an almost periodic function then for
any $\varphi,\ \psi\in{\cal H}$ the following functions are almost periodic:
\ba
t&\rightarrow &W(t)\varphi\no\\
t&\rightarrow &\left(\psi,\ W(t)\varphi\right)\no\\
t&\rightarrow &\left(W(t)\psi,\ W(t)\varphi\right),\no
\ea
where $(\cdot,\cdot)$ denotes the scalar product on ${\cal H}$.
\end{cor}

Another essential result in the theory of almost periodic functions is (see for
example \cite[Property 3, pp.3]{kn:Levitan}):

\begin{theo}\label{th:uct}
Any uniform convergent sequence of almost periodic functions converges towards
an almost periodic function.
\end{theo}

\begin{cor}\label{co:l1conv}
If  $\{\mu_j\}_{j\in\Z}\subset\R$ and $\{\beta_j\}_{j\in\Z}\subseteq X$ satisfies
$\sum_{j\in\Z}\|\beta_j\|<\infty$, then 
$$\sum_{j\in\Z}e^{i\mu_jt}\beta_j$$ is an $X-$ valued almost periodic function
 of $t$.
\end{cor}

\nit\un{Proof:} According to Weierstrass's Criterion the series $\sum_{j\in\Z}e^{i\mu_jt}\beta_j$
is uniformly convergent on $\R$.

By Corollary \ref{co:sum} and Example 1 the partial sums of the above series are
almost periodic. The result follows now from Theorem \ref{th:uct}. \ \ \ \ \ []

We continue with the harmonic analysis results for almost periodic functions.

\begin{theo}(Mean Value)\label{th:meanvalue} 
If $f:\R\rightarrow X$ is almost periodic then the following limit exists and
it is approached uniformly with respect to $a\in\R$:
$$\lim_{t\rightarrow\infty}{1\over t}\int_a^{a+t}f(s)ds=M(f)\in X.$$
Moreover, whenever 
$$t\ge\frac{4\sup_{s\in\R}\|f(s)\|L(\ve/2,f)}{\ve}$$
we have
$$\|M(f)-{1\over t}\int_a^{a+t}f(s)ds\|\le\ve ,$$
for all $a\in\R$.
\end{theo}

The proof of the mean Value Theorem in this form can be found in \cite[pp.
39-44]{kn:Bohr}. Note that although Bohr's book consider only complex valued
almost periodic functions the proof can be carried on to Banach space valued
functions by simply replacing the modulus by the norm and the Lebesque's
integral for complex valued functions by the Bochner's integral.

The results of the next theorem are presented in \cite[Chapter 2]{kn:Levitan}.

\begin{theo}(Fundamental theorem)\label{th:fund}
If $f,\ g:\R\rightarrow X$ are almost periodic then:

\nit (a) for any $\mu\in\R$, 
$$\lim_{t\rightarrow\infty}{1\over t}\int_0^t f(s)e^{-i\mu s}ds=a(\mu,f)$$
exists and is non-zero for at most a denumerable set of $\mu$'s; if
$a(\mu,f)\neq 0$ then $a(\mu,f)$ is called a Fourier coefficient for $f$ while
$\mu$ is called a Fourier exponent;

\nit (b) $a(\mu,f)=a(\mu,g)$ for all $\mu\in\R$ if and only if $f\equiv g;$

\nit (c) let $\Lambda (f)=\{\mu:\ a(\mu,f)\neq 0\}$ denote the set of Fourier
exponents for $f$; then there is an ordering on $\Lambda (f),\ \Lambda
(f)=\{\mu_1,\ \mu_2,\ldots\}$ independent of the Fourier coefficients, 
such that for any $\ve>0$ there exist the numbers
$N(\ve)\in\N,\ 0\le k_{n,\ve}\le 1,\ n\in\N$ with the property that the
trigonometric polynomial
$$ P_\ve (t)=\sum_{n=1}^{N(\ve)}k_{n,\ve} a(\mu_n,f)e^{i\mu_n t}$$
satisfies
$$\|f(t)-P_\ve (t)\|\le\ve\ \ \ {\rm for\ all}\ t\in\R.$$
Moreover $k_{n,\ve}$ can be choosen such that for any fixed $n,\
\lim_{\ve\searrow 0}k_{n,\ve}=1$.
\end{theo}

In this paper we use a less general result than the above Fundamental
Theorem, namely:

\begin{cor}\label{co:fund}
If $f(t)=\sum_{j\in\Z}e^{i\mu_jt}\beta_j$ where $\{\mu_j\}_{j\in\Z}\subset\R$
and  $\sum_{j\in\Z}\|\beta_j\|<\infty$, then $\Lambda(f)=\{\mu_j\}_{j\in\Z}$, 
$a(\mu_j,f)=\beta_j,\ j\in\Z$, in particular if $\mu_j\neq 0,\ j\in\Z$ then
$M(f)=0$. Moreover, we can arbitrarily order $\Lambda (f)$ and still have that 
for any $\ve>0$ there exists a natural number $N(\ve)$ such that:
$$\|f(t)-\sum_{j=-N(\ve)}^{j=N(\ve)}e^{i\mu_jt}\beta_j\|\le\ve,$$
or, in other words, in this particular case the conclusion in part (c) of the 
Fundamental Theorem is valid even if we have an arbitrary order on $\Lambda
(f)$ and we choose $k_{j,\ve}\equiv 1$.
\end{cor}

\nit\un{Proof:} By the Weierstrass criterion the series
$$f(t)e^{-i\mu t}=\sum_{j\in\Z}e^{i(\mu_j-\mu)t}\beta_j$$
is uniformly convergent on $\R$. So, when we compute $a(\mu,f)$ we can
integrate term by term and therefore use the identities: 
$$\lim_{t\rightarrow\infty}{1\over t}\int_0^t e^{-i\la
s}ds=\left\{\begin{array}{rcl}
             0&{\rm if}&\la\neq 0\\
             1&{\rm if}&\la=0
             \end{array}
      \right.
      $$
to get the first part of the Corollary. The last part is a direct consequence
of the fact that $f$ is an absolute and uniform convergent series.\ \ \ \ \ []
\bigskip


\end{document}